\documentstyle[aps,prb,multicol,epsf]{revtex}
\begin{document}
\title{Establishing a non-Fermi liquid theory for disordered metals near two dimensions}
\author{M.A. Baranov$^{1,2}$, I.S. Burmistrov$^{3}$ and A.M.M. Pruisken$^{2}$}

\address{$^{1}$  Kurchatov Institute, \\
Kurchatov Square 1, 123182 Moscow, Russia}
\address{$^{2}$ Institute for Theoretical Physics, University of Amsterdam, \\
Valckenierstraat 65, 1018XE Amsterdam, The Netherlands}
\address{$^{3}$L.D. Landau Institute for Theoretical Physics, \\
Kosygina str. 2, 117940 Moscow, Russia }
\date{\today}
\maketitle

\begin{abstract}
We consider the Finkelstein action describing a system of spin polarized or spinless 
electrons in $2+2\epsilon$ dimensions, in the presence of disorder as well as the Coulomb
interactions. We extend the renormalization group
analysis of our previous work and evaluate the metal-insulator transition of the electron 
gas to second order in an $\epsilon$ expansion. We obtain the complete scaling behavior of 
physical observables like the conductivity and the specific heat with varying frequency, 
temperature and/or electron density.

We extend the results for the interacting electron gas in 
$2+2\epsilon$ dimensions to include the quantum critical behavior of
the plateau transitions in the quantum Hall regime. Although these transitions
have a very different microscopic origin and are controlled by a topological term
in the action ($\theta$ term), the quantum critical behavior is in
many ways the same in both cases. We show that 
the two independent critical exponents of the quantum Hall plateau transitions, 
previously denoted as $\nu$ and $p$, control not only the scaling behavior of
the conductances $\sigma_{xx}$ and $\sigma_{xy}$ at finite temperatures $T$, 
but also the non-Fermi liquid behavior of the specific heat ($c_v \propto T^p$). 
To extract the numerical values of $\nu$ and $p$ 
it is necessary to extend the experiments on transport to include 
the specific heat of the electron gas.
\end{abstract}

\begin{multicols}{2}

\section{Introduction}

\label{Intro}

The integral quantum Hall regime has traditionally been viewed as a (nearly)
free particle localization problem with interactions playing only a minor
role. \cite{freepart1} Although it is well known that many features of the
experimental data, taken from low mobility heterostructures, \cite
{experiments1} can be explained as the behavior of free particles, a much
sharper formulation of the problem is obtained by considering the quantum
Hall plateau transitions.\cite{freepart} Following the experimental work 
by H.P Wei et al.,\cite{experiments}
these transitions behave in all respects like a disorder driven metal-insulator 
transition that is characterized by two independent critical indices, 
i.e. a {\em localization length} 
exponent $\nu$ and a {\em phase breaking length} exponent $p$.\cite{freepart} 
Whereas transport measurements usually provide an experimental value of only the ratio 
$\kappa = p/2\nu$, it is generally not known how the values 
of $\nu$ and $p$ can be extracted separately. 

Inspite of the fact that one can not proceed without having a microscopic theory
of electron-electron interaction effects,
there is nevertheless a strong empirical believe in the literature \cite{believe} 
which says that the zero
temperature localization length exponent $\nu $ is given precisely by the free 
electron value $\nu =2.3$ as obtained from numerical simulations. \cite{kramer}
The experimental situation has not been sufficiently well understood, \cite{newexpt}
however,
to justify the bold assumption of Fermi liquid behavior. In fact, the progress
that has been made over the last few years in the theory of localization and
interaction effects clearly indicates that Fermi liquid principles do not
exist in general. The Coulomb interaction problem lies in a different
universality class of transport phenomena \cite{eurolett} with a previously unrecognized
symmetry, called ${\cal F}$ invariance. \cite{bps1,bps2,bps3,pbb}
The theory relies in many ways on the approach as initiated by Finkelstein 
\cite{fink} and adapted to the case of the spin polarized or spinless
electrons.\cite{eurolett} By reconciling the Finkelstein theory with the
topological concept of an instanton vacuum \cite{inst} and the Chern Simons
statistical gauge fields,\cite{cs} the foundations have been laid for a
complete renormalization theory that unifies the quantum theory of metals
with that of the abelian quantum Hall states.\cite{pbb}

\subsection{A historical problem}

The unification of the integral and fractional quantum Hall regimes is based
on the assumption that Finkelstein approach \cite{fink,eurolett} is
renormalizable and generates a strong coupling, {\em insulating} phase with
a massgap. However, the traditional analyses of the Finkelstein theory have
actually not provided any garantee that this is indeed so.

Inspite of Finkelstein's pioneering and deep contributions to the field, it
is well known that the conventional momentum shell renormalization schemes
do not facilitate any computations of the quantum theory beyond one loop
order. At the same time, application of the more advanced technique of
dimensional regularization has led to conceptual difficulties with such
aspects like {\em dynamical} scaling.\cite{kb} One can therefore not rule
out the possibility that there are complications, either in the idea of
renormalizability, or in other aspects of the theory such as the Matsubara
frequency technique.

Nothing much has been clarified, however, by repeating similar kinds of
analyses in a different formalism, like the Keldish technique.\cite{ka,cln}
What has been lacking all along is the understanding of a fundamental
principle that has prevented the Finkelstein approach from becoming a fully
fledged field theory for localization and interaction effects.

\subsection{${\cal F}$ invariance}

In our previous work \cite{bps1} we have shown that the Finkelstein action
has an exact symmetry (${\cal F}$ invariance) that is intimidly related to
the electrodynamic $U(1)$ gauge invariance of the theory. ${\cal F}$
invariance is the basic mechanism that protects the renormalization of the
problem with infinitely ranged interaction potentials such as the Coulomb
potential. Moreover, it has turned out that the infrared behavior of
physical observables can only be extracted from ${\cal F}$ invariant
quantities and correlations, and these include the linear response to
external potentials. Arbitrary renormalization group schemes break
the ${\cal F}$ invariance of the action and this generally complicates the 
attempt to
obtain the temperature and/or frequency dependence of physical quantities
such as the conductivity and specific heat.

Quantum Hall physics is in many ways a unique laboratory for investigating
and exploring the various different consequences of ${\cal F}$ invariance.
For example, one of the longstanding questions in the field is whether and
how the theory {\em dynamically} generates the {\em exact} quantization of
the Hall conductance. Important progress has been made recently by
demonstrating that the instanton vacuum, on the strong coupling side of the
problem, generally displays massless excitations at the edge of the system. 
\cite{pbv} These massless edge excitations are identical to those described
by the more familiar theory of chiral edge bosons. \cite{bps3} Our theory of
massless edge excitations implies that the concept of ${\cal F}$ invariance
retains its fundamental significance all the way down to the regime of
strong coupling.

\subsection{Outline of this paper}

In this paper we put the concept of ${\cal F}$ invariance at work and
evaluate the renormalization behavior of the Finkelstein theory at a two
loop level. As shown in our previous papers, \cite{bps2} the
technique of dimensional regularization is a unique procedure, not only for
the computation of critical indices, but also for extracting the dynamical
scaling functions. In fact, the metal-insulator transition in $2+2\epsilon$
spatial dimensions is the only place in the theory where the temperature
and/or frequency dependence of physical observables can be obtained
explicitly. This motivates us to further investigate the problem in
$2+2\epsilon $ dimensions and use it as a stage setting for the much more
complex problem of the quantum Hall plateau transitions. 

The final results of this paper are remarkably similar to those of the more
familiar classical Heisenberg ferromagnet.\cite{bhz} For example, unlike
the free electron gas, the Coulomb interaction problem displays a
conventional phase transition (metal-insulator transition) in $2+2\epsilon$
dimensions with an ordinary order parameter. The theory is therefore quite
different from that of free electrons which has a different dimensionality
and displays, as is well known, anomalous or multifractal density
fluctuations near criticality.\cite{weg1}

It is important to bear in mind, however, that the analogy with the
Heisenberg model is rather formal and it fails on many other fronts. For
example, the classification of critical operators is very different from
what one is used to, in ordinary sigma models. Moreover, the Feynman
diagrams of the Finkelstein theory are more complex, involving internal
frequency sums which indicate that the theory effectively exists in $2+1$
space-time dimensions, rather than in two spatial dimensions alone. The
complexity of ${\cal F}$ invariant systems is furthermore illustrated by the
lack of such principles like Griffith analyticity that 
facilitates a discussion of the symmetric phase in conventional sigma models.\cite{bzl}
In a subsequent paper we shall address the
strong coupling insulating phase of the electron gas and show that the
dynamics is distinctly different from that of the Goldstone (metallic) phase
and controlled by different operators in the theory.\cite{pb01}

This paper is organized as follows. After introducing the formalism
(Section II) we embark on the details of the two loop contributions
to the conductivity in Section III. As in our earlier work, we employ an $%
{\cal F}$-invariance-breaking parameter $\alpha $ to regularize the infinite
sums over frequency. This methodology actually provides numerous self
consistency checks and a major part of the computation consists of finding
the ways in which the various singular contributions in $\alpha $ cancel
each other. The actual computation of the diagrams is described in the
Appendices which contain the list of the momentum and frequency integrals
that are used in the text. In tables I and II we summarize how the
different singular contributions in $\alpha $ cancel each other. Table III
lists the various finite contributions to the pole term in $\epsilon $. The
final result for the $\beta $ function is given by Eqs. (\ref{z1})--(\ref{pf}).

In Section IV we summarize the consequences for scaling. We extend the 
discussion to include the plateau transitions in the 
quantum Hall regime in Section IVC. We briefly address several new advancements, both from
theoretical and experimental sides, that seem to have general consequences for the
quantum theory of conductances. Finally, we show how the results of this
paper can be used in the problem of critical exponents $p$ and $\nu$.

We end this paper with a conclusion (Section V).

\section{Effective parameters}

\subsection{Introduction}

The theory for spinless electrons involves unitary matrix field variables $%
Q_{nm}^{\alpha \beta }$ where the superscripts ${\alpha \beta }$ are the 
{\em replica} indices, the subscripts $n$, $m$ denote the {\em Matsubara
frequency} indices. The $Q$ fields obey the nonlinear constraint $Q^{2}=1$
and we are interested in the following action
\begin{eqnarray}
S[Q,A] &=&-\frac{\sigma _{0}}{8}\int_{x}\left( \mbox{tr}[{\vec{D}}%
,Q]^{2}+2h_{0}^{2}\mbox{tr}\Lambda Q\right) +  \nonumber \\
z_{0}\pi &T& \int_{x}\left( {}\right. \sum_{\alpha n}c_{0}\mbox{tr}%
I_{n}^{\alpha }Q\mbox{tr}I_{-n}^{\alpha }Q+4\mbox{tr}\eta Q-6\mbox{tr}\eta
\Lambda \left. {}\right) .  \label{S}
\end{eqnarray}
The explanation of the symbols is as follows. The parameter $\sigma _{0}$
plays the role of {\em conductivity} of the electron gas, $z_{0}$ is the
so-called {\em singlet interaction amplitude} and $T$ stands for the
temperature. The parameter $c_{0}=1-\alpha $ is such that the theory
interpolates between the Coulomb case ($\alpha =0$) and the free particle
case ($\alpha =1$). Here, the quantity $\alpha $ breaks the ${\cal F}$
invariance of the theory and we shall eventually be interested in the limit
where $\alpha $ goes to zero. For a detailed exposure to the meaning of 
${\cal F}$ invariance we refer the reader to the original papers.\cite
{bps1,bps2}

We generally need the definition of two more diagonal matrices $\Lambda$ and 
$\eta$, and one more off-diagonal matrix $I^{\alpha}_{n}$. They are given by

\begin{eqnarray*}
\Lambda _{nm}^{\alpha \beta } &=&{\rm sign}(n)\delta ^{\alpha \beta }\delta
_{nm}, \\
\eta _{nm}^{\alpha \beta } &=&n\delta ^{\alpha \beta }\delta _{nm}, \\
\left( I_{n}^{\alpha }\right) _{kl}^{\beta \gamma } &=&\delta ^{\alpha \beta
}\delta ^{\alpha \gamma }\delta _{n,k-l}.
\end{eqnarray*}
Here, $\eta $, being multiplied by $2\pi T$, represents the Matsubara
frequencies in matrix language. The $I_{n}^{\alpha }$ are shifted diagonals
in frequency space and they generally represent the generators of the $U(1)$
gauge transformations.

The term proportional to $h_{0}^{2}$ is not a part of the theory but we
shall use it later on as a convenient infrared regulator of the theory.
Finally, the $\vec{D}$ are covariant derivatives 
\[
D_{a}=\nabla _{a}-i\widehat{A}_{a}, 
\]
where 
\[
\widehat{A}_{a}=\sum_{\alpha ,n}\left( A_{a}\right) _{n}^{\alpha
}I_{n}^{\alpha }, 
\]
and $\left( A_{a}\right) _{n}^{\alpha }$ is the Fourier transform of the
homogeneous external vector potential $A_{a}^{\alpha }(\tau )$: $%
A_{a}^{\alpha }(\tau )=\sum_{n}\left( A_{a}\right) _{n}^{\alpha }\exp
(-i\omega _{n}\tau )$, $\omega _{n}=2\pi Tn$ is the Matsubara frequency.

\subsection{Linear response}

The ''effective'' action for the external vector potential is defined
according to 
\begin{equation}
\exp S_{eff}[A]=\int DQ\exp S[Q,A].
\end{equation}
The quadratic part can generally be written as 
\begin{equation}
S_{eff}[A]=\int\limits_{x}\sum_{\alpha ,n>0}\sigma ^{^{\prime
}}(n)n(A_{a})_{n}^{\alpha }(A_{a})_{-n}^{\alpha }.
\end{equation}

The quantity $\sigma ^{^{\prime }}(n)$ is the true conductivity of the
electron gas, and in terms of the $Q$ matrix fields the following Kubo like
expression can be obtained

\begin{equation}
\sigma ^{^{\prime }}(n)=\langle O_{1}\rangle +\langle O_{2}\rangle ,
\label{cond1}
\end{equation}
where 
\begin{equation}
O_{1}=-\frac{\sigma _{0}}{4n}\mbox{tr}[I_{n}^{\alpha },Q(x)][I_{-n}^{\alpha
},Q(x)]
\end{equation}
and

\begin{equation}
O_{2}=\frac{\sigma _{0}^{2}}{16nd}\int_{x-x^{\prime }}\mbox{tr}%
[I_{n}^{\alpha },Q(x)]\nabla Q(x)\mbox{tr}[I_{-n}^{\alpha },Q(x^{^{\prime
}})]\nabla Q(x^{^{\prime }}).  \label{cond2}
\end{equation}
Here the expectations are with respect to the theory without the vector
potentials.

\subsection{The $h_0$ field}

Although we are interested, strictly speaking, in evaluating $\sigma
^{^{\prime }}(n)$ with varying values of external frequencies $\omega _{n}$
and temperature, the computation simplifies dramatically if we put these
parameters equal to zero in the end and work with a finite value of the $%
h_{0}$ field instead. This procedure has been analyzed in exhaustive detail
in our previous work and, in what follows, we shall greatly benifit from the
technical advantages that make the two-loop analysis of the conductivity
possible. We shall return to finite frequency and temperature problem in the
end of this paper (Sections IV).

The infrared regularization by the $h_{0}$ field relies on the following
statement 
\begin{equation}
\sigma _{0}h_{0}^{2}\langle Q({\vec{x}})\rangle =\sigma ^{^{\prime
}}h^{^{\prime }2}\Lambda ,
\end{equation}
which says that there is an effective mass $h^{^{\prime }}$ in the problem
that is being induced by the presence of the $h_{0}$ field. It is very well
known that, since the quantity $\langle Q({\vec{x}})\rangle $ is not a gauge
invariant object, the definition of the $h^{^{\prime }}$ field is singular
as $\alpha $ goes to zero and the theory is generally not renormalizable.
However, the effective parameter $\sigma ^{^{\prime }}$ is truly defined in
terms of the effective mass $h^{^{\prime }}$ rather than the bare parameter $%
h_{0}$. Hence, all the non-renormalizable singularities are removed from the
theory, provided we express $\sigma ^{^{\prime }}$ in terms of the $%
h^{^{\prime }}$ rather than the $h_{0}$. We shall show that the ultraviolet
singularities of the theory can be extracted directly from the final result
for $\sigma ^{^{\prime }}(h^{^{\prime }})$. On the other hand, we can make
use of our previous results \cite{bps2} and express the final answer in
terms of frequencies and temperature, rather than the mass $h^{^{\prime }}$.

\section{Computation of conductivity in $2+2\epsilon$ dimensions}

\label{PEC}

\subsection{Introduction}

To define a theory for perturbative expansions we use the following
parametrization 
\begin{equation}
Q=\left( 
\begin{array}{cc}
\sqrt{1-qq^{\dagger }} & q^{\dagger } \\ 
q & -\sqrt{1-q^{\dagger }q}
\end{array}
\right) .
\end{equation}
The action can be written as an infinite series in the independent fields $%
q_{n_{1}n_{2}}^{\alpha \beta }$ and $[q^{\dagger }]_{n_{4}n_{3}}^{\alpha
\beta }$. We use the convention that Matsibara indices with odd subscripts: $%
n_{1},n_{3},...$, run over non-negative integers, whereas those with even
subscripts: $n_{2},n_{4},...$, run over negative integers. The propagators
can be written in the form ~\cite{kb,bps2} 
\begin{eqnarray}
\langle q_{n_{1}n_{2}}^{\alpha \beta }(p)[q^{\dagger }]_{n_{4}n_{3}}^{\gamma
\delta }(-p)\rangle &=&\frac{4}{\sigma _{0}}\delta ^{\alpha \delta }\delta
^{\beta \gamma }\delta _{n_{12},n_{34}}D_{p}(n_{12})  \nonumber \\
\left( {}\right. \delta _{n_{1}n_{3}} &+&\delta ^{\alpha \beta }\kappa
^{2}z_{0}c_{0}D_{p}^{c}(n_{12})\left. {}\right) ,
\end{eqnarray}
where 
\begin{equation}
\lbrack D_{p}(n_{12})]^{-1}=p^{2}+h_{0}^{2}+\kappa ^{2}z_{0}n_{12},
\end{equation}
\begin{equation}
\lbrack D_{p}^{c}(n_{12})]^{-1}=p^{2}+h_{0}^{2}+\alpha \kappa
^{2}z_{0}n_{12},
\end{equation}
\begin{equation}
\kappa ^{2}=\frac{8\pi T}{\sigma _{0}}.
\end{equation}
Here we use the notation $n_{12}=n_{1}-n_{2}$ .

The expression for the DC conductivity is known to one loop order~\cite{bps2}
\begin{equation}
\sigma _{one}^{^{\prime }}=\sigma _{0}+\frac{4\Omega _{d}h_{0}^{2\epsilon }}{%
\epsilon }\,\,,\,\,\Omega _{d}=\frac{S_{d}}{2(2\pi )^{d}},  \label{one}
\end{equation}
where $S_{d}=2\pi ^{d/2}/\Gamma (d/2)$ is the surface of a $d$ dimensional
unit sphere.

\subsection{ The two-loop theory}

To proceed we need the following terms obtained by expanding the action
(2.1) in terms of $q$ and $q^{\dagger }$ fields: 
\begin{eqnarray}
S_{int}^{(3)}=-\frac{a\sigma _{0}}{8}\int\limits_{x}\sum\limits_{\beta
,m>0}\left\{ {}\right. &\mbox{tr}&I_{m}^{\beta }q^{\dagger }\,\mbox{tr}%
I_{-m}^{\beta }[q,q^{\dagger }]+  \nonumber \\
&\mbox{tr}&I_{-m}^{\beta }q\,\mbox{tr}I_{m}^{\beta }[q,q^{\dagger }]\left.
{}\right\} ,
\end{eqnarray}
\begin{eqnarray}
S_{int}^{(4)}=\frac{a\sigma _{0}}{16}\int\limits_{x}\left\{ {}\right.
\sum\limits_{\beta ,m>0} &\mbox{tr}&I_{-m}^{\beta }[q,q^{\dagger }]\,%
\mbox{tr}I_{m}^{\beta }[q,q^{\dagger }]+  \nonumber \\
2\sum\limits_{\beta }( &\mbox{tr}&I_{0}^{\beta }[q,q^{\dagger }])^{2}\left.
{}\right\} ,
\end{eqnarray}
\begin{eqnarray}
S_{0}^{(4)} &=&\frac{\sigma _{0}}{32}\int\limits_{p}\delta ({\bf p}_{1}+{\bf %
p}_{2}+{\bf p}_{3}+{\bf p}_{4})\sum\limits_{n_{1}n_{2}n_{3}n_{4}}^{\beta
\gamma \delta \mu }  \nonumber \\
&\times &q_{n_{1}n_{2}}^{\beta \gamma }(p_{1})(q^{\dagger
})_{n_{2}n_{3}}^{\gamma \delta }(p_{2})q_{n_{3}n_{4}}^{\delta \mu
}(p_{3})(q^{\dagger })_{n_{4}n_{1}}^{\mu \beta }(p_{1})  \nonumber \\
&\times &\left\{ {}\right. ({\bf p}_{1}+{\bf p}_{2})\cdot ({\bf p}_{3}+{\bf p}
_{4})+({\bf p}_{2}+{\bf p}_{3})\cdot ({\bf p}_{1}+{\bf p}_{4})  \nonumber \\
&-&\kappa ^{2}z_{0}(n_{12}+n_{34})-2h_{0}^{2}\left. {}\right\} ,
\end{eqnarray}
where we define $a=\kappa ^{2}z_{0}c_{0}$.

In addition, we need the following terms obtained by expanding the
expression for the conductivity, Eq.~(\ref{cond1}), 
\begin{eqnarray}
O_{1}^{(2)}=-\frac{\sigma _{0}}{2} &\mbox{tr}&\left\{ {}\right.
I_{n}^{\alpha }q^{\dagger }I_{-n}^{\alpha }q+I_{-n}^{\alpha }q^{\dagger
}I_{n}^{\alpha }q-  \nonumber \\
&2&(I_{n}^{\alpha }\Lambda I_{-n}^{\alpha }+I_{-n}^{\alpha }\Lambda
I_{n}^{\alpha })[q,q^{\dagger }]\left. {}\right\} ,
\end{eqnarray}
\begin{equation}
O_{1}^{(3)}=\frac{\sigma _{0}}{4}\mbox{tr}\left\{ {}\right. I_{n}^{\alpha
}(q+q^{\dagger })I_{-n}^{\alpha }qq^{\dagger }-I_{-n}^{\alpha }(q+q^{\dagger
})I_{n}^{\alpha }q^{\dagger }q\left. {}\right\} ,
\end{equation}
\begin{eqnarray}
O_{1}^{(4)}=\frac{\sigma _{0}}{16} &\mbox{tr}&\left\{ {}\right.
(I_{n}^{\alpha }\Lambda I_{-n}^{\alpha }+I_{-n}^{\alpha }\Lambda
I_{n}^{\alpha })[qq^{\dagger }q,q^{\dagger }]-  \nonumber \\
&2&I_{n}^{\alpha }[q,q^{\dagger }]I_{-n}^{\alpha }[q,q^{\dagger }]\left.
{}\right\} ,
\end{eqnarray}
\begin{equation}
O_{2}^{(4)}=\frac{\sigma _{0}^{2}}{4d}\int_{x-x^{\prime }}\mbox{tr}%
I_{n}^{\alpha }(q\nabla q^{\dagger }+q^{\dagger }\nabla q)\mbox{tr}%
I_{-n}^{\alpha }(q\nabla q^{\dagger }+q^{\dagger }\nabla q),
\end{equation}
\begin{eqnarray}
O_{2}^{(5)}=\frac{\sigma _{0}^{2}}{8d}\int_{x-x^{\prime }}\left\{ {}\right. &%
\mbox{tr}&I_{n}^{\alpha }(q\nabla q^{\dagger }+q^{\dagger }\nabla q)\mbox{tr}%
I_{-n}^{\alpha }q(\nabla q^{\dagger })q+  \nonumber \\
&\mbox{tr}&I_{-n}^{\alpha }(q\nabla q^{\dagger }+q^{\dagger }\nabla q)%
\mbox{tr}I_{n}^{\alpha }q^{\dagger }(\nabla q)q^{\dagger }\left. {}\right\} ,
\end{eqnarray}
\begin{eqnarray}
O_{2}^{(6)}=\frac{\sigma _{0}^{2}}{16d}\int_{x-x^{\prime }}\left\{ {}\right.
&\mbox{tr}&I_{n}^{\alpha }\Lambda q^{\dagger }(\nabla q)q^{\dagger })%
\mbox{tr}I_{-n}^{\alpha }\Lambda q(\nabla q^{\dagger })q)+  \nonumber \\
&\mbox{tr}&I_{n}^{\alpha }(q\nabla q^{\dagger }+q^{\dagger }\nabla q)\times 
\nonumber \\
&\mbox{tr}&I_{-n}^{\alpha }(qq^{\dagger }\nabla (qq^{\dagger })+q^{\dagger
}q\nabla (q^{\dagger }q))+  \nonumber \\
&\mbox{tr}&I_{-n}^{\alpha }(q\nabla q^{\dagger }+q^{\dagger }\nabla q)\times
\nonumber \\
&\mbox{tr}&I_{n}^{\alpha }(qq^{\dagger }\nabla (qq^{\dagger })+q^{\dagger
}q\nabla (q^{\dagger }q))\left. {}\right\} .
\end{eqnarray}

Next we give the complete list of two loop contributions to the conductivity
as follows 
\begin{eqnarray}
&\sigma &_{two}^{^{\prime }}(n)=  \nonumber \\
&\langle
&O_{1}^{(4)}+O_{1}^{(3)}S_{int}^{(3)}+O_{1}^{(2)}(S_{int}^{(4)}+S_{0}^{(4)}+%
\frac{1}{2}(S_{int}^{(3)})^{2})  \nonumber \\
&+&O_{2}^{(6)}+O_{2}^{(5)}S_{int}^{(3)}+O_{2}^{(4)}(S_{int}^{(4)}+S_{0}^{(4)}+%
\frac{1}{2}(S_{int}^{(3)})^{2})\rangle .  \label{sig}
\end{eqnarray}

The computations of the terms in Eq. (\ref{sig}) are straightforward but
lengthy and tedious. In what follows we present the expressions in terms of
the momentum integrals, frequency sums and propagators $D$, $D^{c}$ for each
term in Eq. (\ref{sig}) separately, along with the final answer. In the
Appendices we give the complete list of integrals and symbols that we shall
make use of here.

\subsection{Computation of contractions}

\subsubsection{$\langle O_{1}^{(4)} \rangle$}

\begin{eqnarray}
&&\frac{2}{\sigma _{0}}(\int\limits_{p}D_{p}(0))^{2}+\frac{2a^{2}}{\sigma
_{0}}(\sum_{m>0}\int\limits_{p}DD_{q}^{c}(m))^{2}  \nonumber \\
&=&\frac{\Omega _{d}^{2}h_{0}^{4\epsilon }}{\sigma _{0}\epsilon ^{2}}(2+2\ln
^{2}\alpha )  \label{start}
\end{eqnarray}
with $DD_{q}^{c}(m)\equiv D_{q}(m)D_{q}^{c}(m)$.

\subsubsection{$\langle O_{1}^{(3)}S^{(3)}_{int}\rangle$}

\begin{eqnarray}
&-&\frac{8a}{\sigma _{0}}\int\limits_{p,q}\left\{ {}\right.
\sum_{k>0}D_{p+q}^{c}(0)D_{q}(k)D_{p}(k)  \nonumber \\
&+&a\sum_{k,m>0}D_{p}^{c}(m)DD_{q}^{c}(k)D_{p+q}(k+m)\left. {}\right\} 
\nonumber \\
&=&\frac{\Omega _{d}^{2}h_{0}^{4\epsilon }}{\sigma _{0}\epsilon }\left(
{}\right. 4S_{0}+4A_{00}^{0}\left. {}\right)  \nonumber \\
&=&\frac{\Omega _{d}^{2}h_{0}^{4\epsilon }}{\sigma _{0}\epsilon ^{2}}\left[
{}\right. -4-4\ln ^{2}\alpha +\epsilon (8+4\zeta (3))\left. {}\right] ,
\end{eqnarray}
where $\zeta (z)$ is the Riemann zeta-function.

\subsubsection{$\langle O_{1}^{(2)}(S^{(4)}_{int}+S^{(4)}_{0}+\frac{1}{2}%
(S^{(3)}_{int})^{2}))\rangle$}

\begin{eqnarray}
&&\frac{4a}{\sigma _{0}}\int\limits_{p,q}\left\{ {}\right.
D_{p+q}^{c}(0)\sum_{k>0}D_{q}(k)D_{p}(k)  \nonumber \\
&+&a\sum_{k,m>0}D_{p}^{c}(m)D_{q}^{c}(k)D_{p+q}^{2}(k+m)  \nonumber \\
&+&a\sum_{k,m>0}(1+amD_{p}^{c}(m))DD_{q}^{c}(k)D_{p+q}^{2}(k+m)\left.
{}\right\}  \nonumber \\
&=&\frac{\Omega _{d}^{2}h_{0}^{4\epsilon }}{\sigma _{0}\epsilon }\left(
{}\right. -2S_{0}-2D_{1}-2T_{01}-2A_{1,0}^{0}\left. {}\right)  \nonumber \\
&=&\frac{\Omega _{d}^{2}h_{0}^{4\epsilon }}{\sigma _{0}\epsilon ^{2}}\left[
{}\right. 2+2\ln ^{2}\alpha -\epsilon (4+2\zeta (3)+2\pi ^{2}/3)\left.
{}\right] .
\end{eqnarray}

\subsubsection{$\langle O_{2}^{(6)}\rangle$}

\begin{eqnarray}
&-&\frac{4}{\sigma _{0}d}\int\limits_{p,q}p^{2}\left\{ {}\right.
D_{p}(0)D_{q}(0)D_{p+q}(0)  \nonumber \\
&-&4a^{2}\sum_{k,m>0}D^{2}D_{p}^{c}(m)\hat{S}_{m}DD_{q}^{c}(k)  \nonumber \\
&-&a^{2}\sum_{k,m>0}\left[ {}\right. D_{p}(k+m)DD_{q}^{c}(m)DD_{p+q}^{c}(k) 
\nonumber \\
&+&2DD_{p}^{c}(k+m)DD_{q}^{c}(m)D_{p+q}(k)\left. {}\right] \left. {}\right\}
\nonumber \\
&=&\frac{\Omega _{d}^{2}h_{0}^{4\epsilon }}{\sigma _{0}\epsilon }\left(
{}\right. S_{1}+4(\frac{2\ln \alpha }{\epsilon }+B_{1})+C_{01}+2C_{00}\left.
{}\right)  \nonumber \\
&=&\frac{\Omega _{d}^{2}h_{0}^{4\epsilon }}{\sigma _{0}\epsilon ^{2}}\left[
{}\right. 16\ln \alpha -2+\epsilon (-4\ln \alpha -\frac{\pi ^{2}}{3}+\frac{%
\pi ^{2}}{2}\ln 2  \nonumber \\
&+&\frac{\pi ^{4}}{12}+\frac{11\zeta (3)}{2}+\frac{\pi ^{2}}{3}\ln ^{2}2-%
\frac{1}{3}\ln ^{4}2-7\zeta (3)\ln 2  \nonumber \\
&-&8Li_{4}(\frac{1}{2}))\left. {}\right] .
\end{eqnarray}
Here $D^{n}D_{q}^{c}(m)\equiv D_{q}^{n}(m)D_{q}^{c}(m)$ and

\begin{equation}
Li_{n}(x)=\sum\limits_{k=1}^{\infty }\frac{x^{k}}{k^{n}}
\end{equation}
is the polylogarithmic function ($Li_{4}(1/2)=0.517...$), and we have
introduced an operator $\hat{S}_{m}$ which acts only on frequency $k$
according to the following rule $\hat{S}_{m}f(k)=f(k)+f(k+m)$.

\subsubsection{$\langle O_{2}^{(5)}S^{(3)}_{int}\rangle$}

\begin{eqnarray}
&&\frac{16a}{\sigma _{0}d}\int\limits_{p,q}{\bf p}\cdot({\bf p}-{\bf q}%
)\sum_{k>0}D_{p+q}^{c}(0)D_{p}^{2}(k)D_{q}(k)  \nonumber \\
&+&\frac{16a^{2}}{\sigma _{0}d}\int\limits_{p,q}p^{2}%
\sum_{k,m>0}D_{p+q}^{c}(m)\left[ {}\right. D_{p}^{2}(k+m)DD_{q}^{c}(k) 
\nonumber \\
&+&D^{2}D_{p}^{c}(k+m)D_{q}(k)\left. {}\right]  \nonumber \\
&-&\frac{16a^{2}}{\sigma _{0}d}\int\limits_{p,q}({\bf p}\cdot {\bf q})\sum_{k,m>0}
\left\{ 
{}\right. DD_{p}^{c}(m)\hat{T}_{m}DD_{p+q}^{c}(k)D_{q}(k+m)  \nonumber \\
&+&D_{p+q}^{c}(m)D^{2}D_{p}^{c}(k+m)D_{q}(k+2m)\left. {}\right\}  \nonumber
\\
&=&\frac{\Omega _{d}^{2}h_{0}^{4\epsilon }}{\sigma _{0}\epsilon }\left(
{}\right. -4S_{00}-4A_{01}^{1}-4H_{0}-4C_{0}-4A_{0}\left. {}\right) 
\nonumber \\
&=&\frac{\Omega _{d}^{2}h_{0}^{4\epsilon }}{\sigma _{0}\epsilon ^{2}}\left[
{}\right. -8\ln \alpha +4+\epsilon (4\ln ^{2}\alpha +20\ln \alpha  \nonumber
\\
&-&12+4\zeta (3)+4\pi ^{2}/3-4A_{0}+4C_{0}^{^{\prime }})\left. {}\right] ,
\end{eqnarray}
Where we have introduced yet another operator $\hat{T}_{m}$ which acts only
on frequency $k$ but now according to the rule $\hat{T}_{m}f(k)=f(k)-f(k+m)$.

\subsubsection{$\langle O_{2}^{(4)}S^{(4)}_{0}\rangle$}

\begin{eqnarray}
&&\frac{8a^{2}}{\sigma _{0}d}\int\limits_{p,q}p^{2}\sum_{k,m>0}  \nonumber \\
&\left\{ {}\right. &3D^{3}D_{p}^{c}(m)\hat{S}%
_{m}D_{q}^{c}(k)+3D^{2}D_{p}^{c}(m)\hat{S}_{m}DD_{q}^{c}(k)  \nonumber \\
&+&2akD^{2}[D_{p}^{c}]^{2}(m)\hat{T}_{m}[D_{p}(m)D_{q}^{c}(k)+DD_{q}^{c}(k)]%
\left. {}\right\}  \nonumber \\
&=&\frac{\Omega _{d}^{2}h_{0}^{4\epsilon }}{\sigma _{0}\epsilon }\left(
{}\right. -3T_{10}^{0}-3T_{11}^{0}-\frac{12\ln \alpha }{\epsilon }%
-6B_{1}-2T_{20}^{0}  \nonumber \\
&+&2T_{21}^{0}-4T_{10}^{1}+4B_{2}\left. {}\right)  \nonumber \\
&=&\frac{\Omega _{d}^{2}h_{0}^{4\epsilon }}{\sigma _{0}\epsilon ^{2}}\left[
{}\right. 4(\ln \alpha -1)^{2}-2 \\
&+&\epsilon (\frac{2}{\alpha }-2\ln ^{2}\alpha +\ln \alpha +44/3)\left.
{}\right] .
\end{eqnarray}

\subsubsection{$\langle O_{2}^{(4)}(S^{(4)}_{int}+\frac{1}{2}%
(S^{(3)}_{int})^{2})\rangle$}

\begin{eqnarray}
&&\frac{16a^{2}}{\sigma _{0}d}\int\limits_{p,q}({\bf p}\cdot {\bf q}
)\sum_{k>0}kD_{p+q}(0)D_{p}^{2}(k)D_{q}^{2}(k)  \nonumber \\
&-&\frac{16a}{\sigma _{0}d}\int\limits_{p,q}p^{2}\sum_{k>0}\left[ {}\right.
2akD_{p+q}^{c}(0)D_{p}^{3}(k)D_{q}(k)-D_{p}^{3}(k)D_{q}(k)\left. {}\right] 
\nonumber \\
&+&\frac{16a}{\sigma _{0}d}\int\limits_{p,q}({\bf pq})\sum_{k,m>0}\left\{
{}\right. 2(1+amD_{q}(k))  \nonumber \\
&\times &D_{p+q}^{c}(m)D_{p}^{2}(k+m)DD_{q}^{c}(k)  \nonumber \\
&-&DD_{q}^{c}(k)DD_{p}^{c}(m)D_{p+q}(k+m)\left. {}\right\}  \nonumber \\
&-&\frac{16a^{2}}{\sigma _{0}d}\int\limits_{p,q}p^{2}\sum_{k,m>0}\left\{
{}\right. (1+amD_{p+q}^{c}(m))  \nonumber \\
&\times &\left[ {}\right. (2+\hat{T}_{m}+ak\hat{T}%
_{m}D_{p}^{c}(k))D^{3}D_{p}^{c}(k)D_{q}(k+m)  \nonumber \\
&+&\frac{1}{2}D_{q}(k)D_{p}^{3}(k+m)(3D_{q}^{c}(k)+D_{p}^{c}(k+m))\left.
{}\right]  \nonumber \\
&+&\frac{3}{2}D_{q}^{c}(m)D_{p+q}^{c}(k)D_{p}^{3}(k+m)  \nonumber \\
&+&(1+\hat{T}_{m}+2ak\hat{T}%
_{m}D_{p}^{c}(k))D_{p+q}^{c}(m)D^{2}D_{p}^{c}(k)D_{q}(k+m)  \nonumber \\
&+&ak\hat{T}_{m}D[D_{p}^{c}(m)]^{2}DD_{q}^{c}(k)D_{p+q}(k+m)\left. {}\right\}
\nonumber \\
&=&\frac{\Omega _{d}^{2}h_{0}^{4\epsilon }}{\sigma _{0}\epsilon }\left(
{}\right. 4S_{11}+4S_{01}+\frac{2}{\epsilon }+8A_{01}+8A_{11}-4C_{11} 
\nonumber \\
&+&4T_{02}+4A_{10}+2T_{12}+2A_{1}+3T_{01}+3A_{11}^{1}+\alpha T_{10}^{0} 
\nonumber \\
&-&T_{02}+H_{1}+3D_{2}+4C_{1}+8A_{2}+8A_{00}-4A_{3}\left. {}\right) 
\nonumber \\
&=&\frac{\Omega _{d}^{2}h_{0}^{4\epsilon }}{\sigma _{0}\epsilon ^{2}}\left[
{}\right. -4\ln ^{2}\alpha -4+\epsilon (-\frac{2}{\alpha }-2\ln ^{2}\alpha 
\nonumber \\
&-&25\ln \alpha +55/2-2\zeta (3)-\frac{8}{3}\pi ^{2}+12\ln ^{2}2  \nonumber
\\
&-&44\ln 2-4C_{0}^{^{\prime }}+4A_{0}+16G-8Li_{2}(\frac{1}{2})\left.
{}\right] ,  \label{end}
\end{eqnarray}
where $G=0.915...$ denotes the Catalan constant.

\subsection{Results of the computations}

We proceed by presenting the final answer for all the pole terms in $%
\epsilon $. By putting the external frequency equal to zero and in the limit 
$\alpha \rightarrow 0$ we obtain 
\begin{equation}
\sigma _{two}^{^{\prime }}(0)=\frac{\Omega _{d}^{2}h_{0}^{4\epsilon }}{%
\sigma _{0}\epsilon }\left( A-8(2+\ln \alpha )\right) . 
\label{res}
\end{equation}
Here, the $A$ stands for all the terms that are finite in $\alpha $. The
complete list is as follows 
\begin{eqnarray}
A &=&50+\frac{1}{6}-3\pi ^{2}+\frac{19}{2}\zeta (3)+16\ln ^{2}2  \nonumber \\
&-&44\ln 2+\frac{\pi ^{2}}{2}\ln 2+16G+\frac{\pi ^{4}}{12}+\frac{\pi ^{2}}{3}%
\ln ^{2}2  \nonumber \\
&-&\frac{1}{3}\ln ^{4}2-7\zeta (3)\ln 2-8Li_{4}(\frac{1}{2})  \nonumber \\
&\approx &1.64.
\end{eqnarray}

Before Eq. (\ref{res}) is obtained, one has to deal with a host of other
contributions that are more singular in $\alpha $ and/or $\epsilon $. These
more singular contributions all cancel one another in the end, however.
There are in total six different types of contributions that are more
singular than the simple pole term $1/\epsilon $. In Tables I and II we list
these terms, show where they come from and how they sum up to zero. There is
one exception, namely the terms proportional to $\ln (\alpha )/\epsilon $,
and their contribution is written in Eq. (\ref{res}). However, these terms
are absorbed in the definition of an ''effective'' $h^{\prime }$ field. More
specifically, from the two-loop computation of the singlet amplitude $z$ we
know that the effective $h,$ field is given by \cite{bps2} 
\begin{equation}
h_{0}^{2}\to h^{\prime 2}=h_{0}^{2}\left( 1-\frac{2+\ln \alpha }{2}\frac{%
h_{0}^{2\epsilon }t_{0}}{\epsilon }\right) .
\end{equation}
Using this result, as well as Eqs. (\ref{one}) and (\ref{res}), we can write
the total answer for the conductivity as follows 
\begin{equation}
\sigma ^{^{\prime }}=\sigma _{0}\left( {}\right. 1+\frac{h^{\prime 2\epsilon
}t_{0}}{\epsilon }+A\frac{h^{\prime 4\epsilon }t_{0}^{2}}{\epsilon }\left.
{}\right) .  \label{sigma}
\end{equation}
Here we have written $t_{0}=4\Omega _{d}/\sigma _{0}$. Eq. (\ref{sigma}) no
longer contains $\alpha $ and is therefore the desired result.

\subsection{$\beta$ and $\gamma$ functions}

Recall that $h$ is just the effective mass in the problem and we can replace
it by the effective mass that is being induced by working with finite
external frequencies, or finite temperatures. However, we can use Eq. (\ref
{sigma}) directly for extracting the renormalization constant $Z_{1}$ for
the $t$ field. Introducing the renormalized fields $t$ and $z$ as usual 
\begin{equation}
t_{0}=\mu ^{-2\epsilon }tZ_{1}(t),\;\;z_{0}=zZ_{2}(t),
\end{equation}
then, following the scheme of minimal subtraction, we obtain 
\begin{eqnarray}
Z_{1} &=&1+\frac{t}{\epsilon }+\frac{t^{2}}{\epsilon ^{2}}(1+\epsilon A)
\label{z1} \\
Z_{2} &=&1-\frac{t}{2\epsilon }-\frac{t^{2}}{4\epsilon ^{2}}\left( \frac{1}{2%
}+\epsilon (\frac{\pi ^{2}}{6}+2)\right) .  \label{z2}
\end{eqnarray}
Here, we have listed also the result for $Z_{2}$ that was obtained in Ref. 10. 
The $\beta $ and $\gamma $ functions are defined by 
\begin{equation}
\beta =\frac{dt}{d\ln \mu }=\frac{2\epsilon t}{1+td\ln Z_{1}/dt},
\label{genbeta}
\end{equation}
\begin{equation}
\gamma =-\frac{d\ln z}{d\ln \mu }=\beta \frac{d\ln Z_{2}}{dt},
\label{gengamma}
\end{equation}
and the final answer can be written as 
\begin{equation}
\beta =2\epsilon t-2t^{2}-4At^{3}\,\,,\,\,\gamma =-t-(\frac{\pi ^{2}}{6}%
+3)t^{2}.  \label{pf}
\end{equation}

\end{multicols}


\begin{table}[tbp]
\begin{tabular}{|c|c||c|c|c|c|c|c|}
\hline
&  &  &  &  &  &  &  \\ 
Contractions & Diagrams & $\frac{1}{\epsilon\alpha}$ & $\frac{\log^{2}\alpha%
}{\epsilon^{2}}$ & $\frac{\log \alpha}{\epsilon^{2}}$ & $\frac{\log^{2}\alpha%
}{\epsilon}$ & $\frac{\log \alpha}{\epsilon}$ & $\frac{1}{\epsilon^{2}}$ \\ 
&  &  &  &  &  &  &  \\ \hline\hline
&  &  &  &  &  &  &  \\ 
$\langle O_{1}^{(4)} \rangle $ & \epsfxsize = 40pt \epsfysize = 20pt %
\epsffile{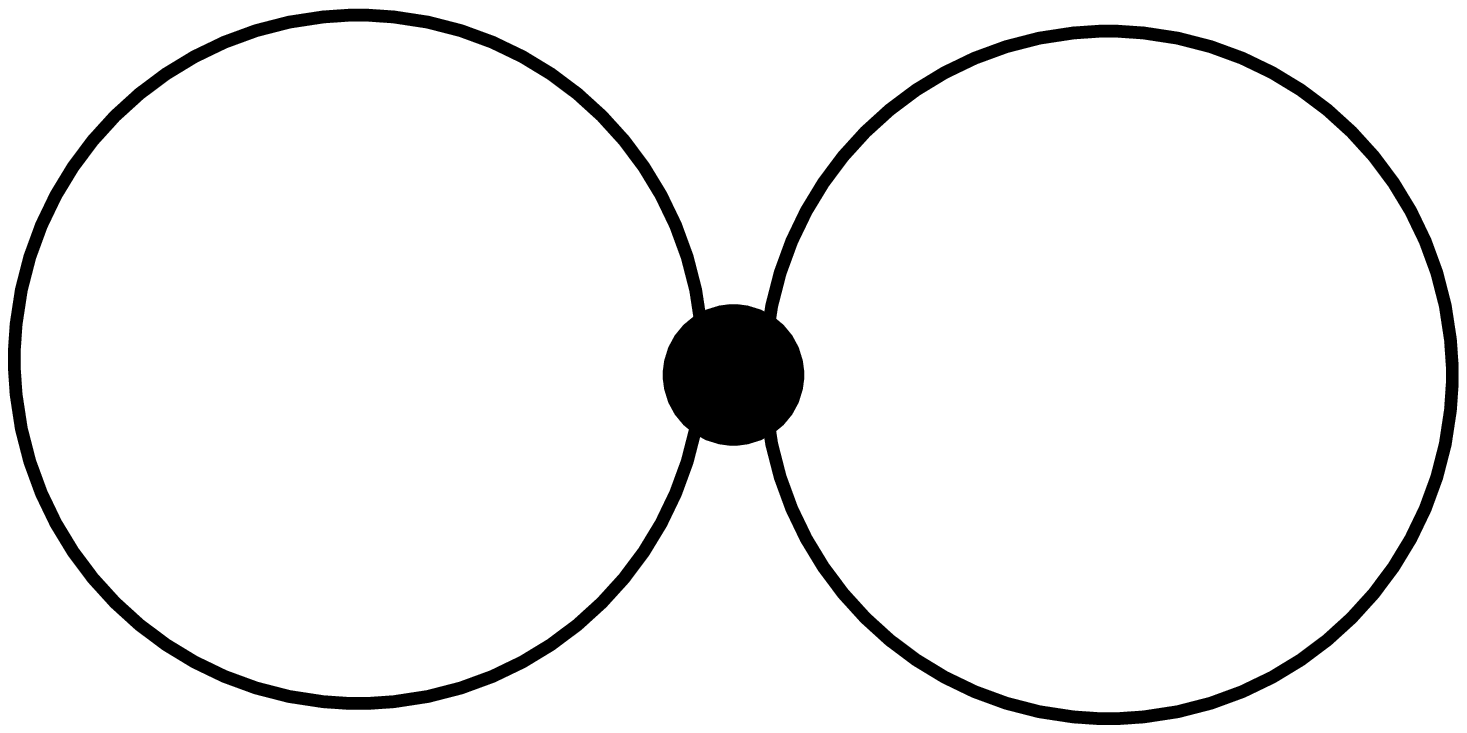} &  & 2 &  &  &  & 2 \\ 
&  &  &  &  &  &  &  \\ \hline
&  &  &  &  &  &  &  \\ 
$\langle O_{1}^{(3)} S^{(3)}_{int} \rangle $ & \epsfxsize = 40pt \epsfysize %
= 20pt \epsffile{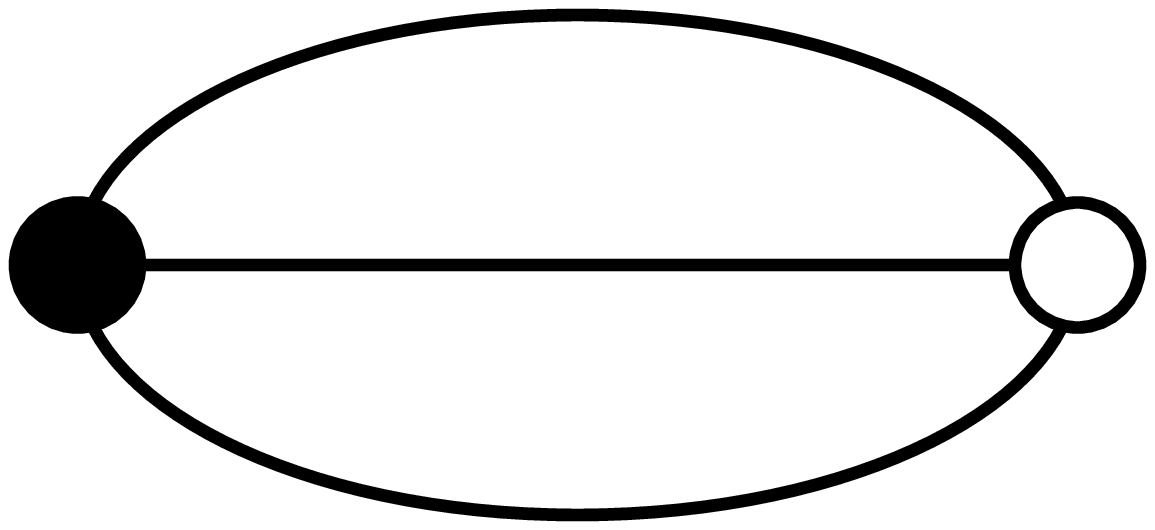} &  & -4 &  &  &  & -4 \\ 
&  &  &  &  &  &  &  \\ \hline
&  &  &  &  &  &  &  \\ 
$\langle O_{1}^{(2)}( S^{(4)}_{int} + S^{(4)}_{0} +\frac{1}{2} (
S^{(3)}_{int})^{2}) \rangle $ & \epsfxsize = 40pt \epsfysize = 20pt %
\epsffile{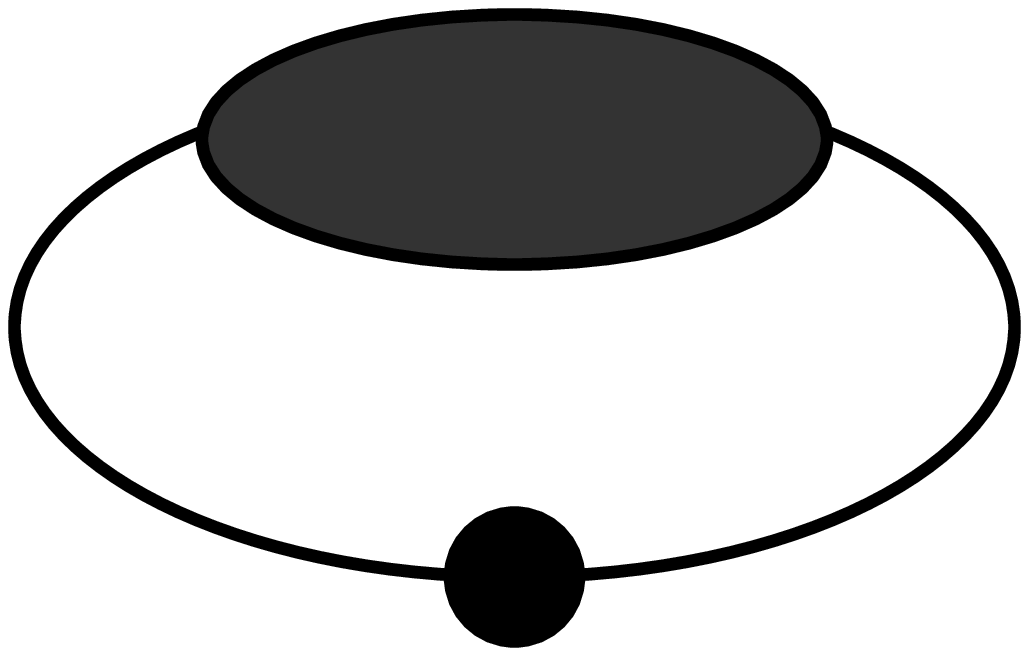} &  & 2 &  &  &  & 2 \\ 
&  &  &  &  &  &  &  \\ \hline\hline
&  &  &  &  &  &  &  \\ 
Total &  & 0 & 0 & 0 & 0 & 0 & 0 \\ 
&  &  &  &  &  &  &  \\ \hline
\end{tabular}
\par
\begin{figure}[tbp]
\end{figure}
\caption{ The second-loop contributions to the $O_{1} $ term in the
effective conductivity. The $\alpha$-dependent and $1/\epsilon^{2}$
contributions. A black solid dot denotes the vertex in $O_{1} $ term, a
white solid dot denotes the vertex in $S $ terms, and }
\begin{figure}
\epsfxsize = 200pt
\epsfysize = 30pt
\epsffile{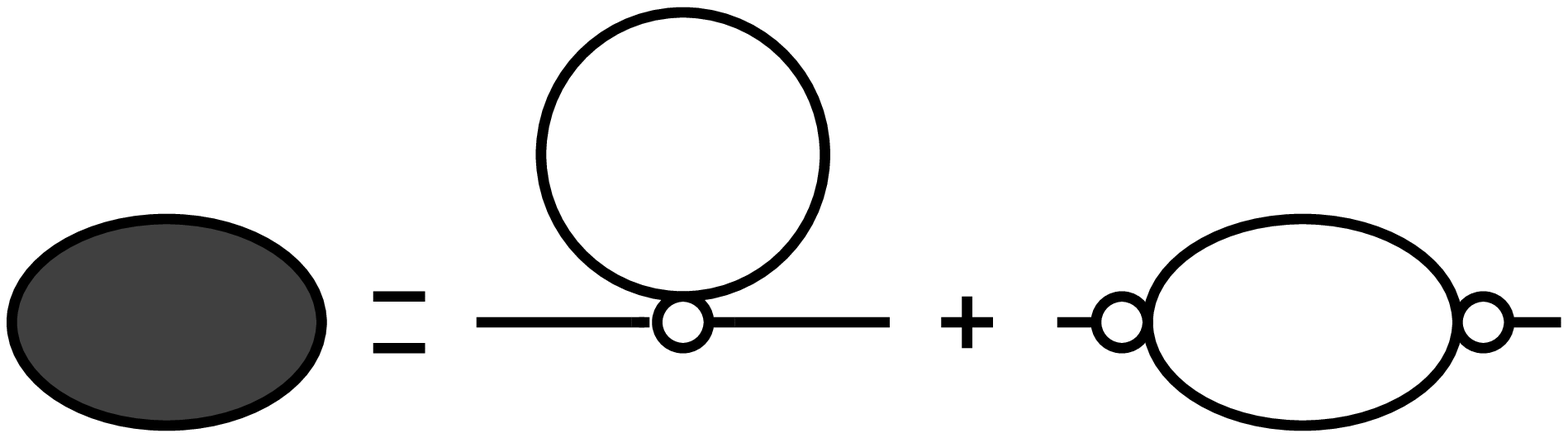}
\end{figure}
\end{table}


\begin{table}[tbp]
\begin{tabular}{|c|c||c|c|c|c|c|c|}
\hline
&  &  &  &  &  &  &  \\ 
Contractions & Diagrams & $\frac{1}{\epsilon \alpha } $ & $\frac{%
log^{2}\alpha}{\epsilon^{2}} $ & $\frac{log\alpha}{\epsilon^{2}} $ & $\frac{%
log^{2}\alpha}{\epsilon} $ & $\frac{log \alpha}{\epsilon} $ & $\frac{1}{%
\epsilon^{2}}$ \\ 
&  &  &  &  &  &  &  \\ \hline\hline
&  &  &  &  &  &  &  \\ 
$\langle O^{(6)}_{2} \rangle $ & \epsfxsize = 40pt \epsfysize = 20pt %
\epsffile{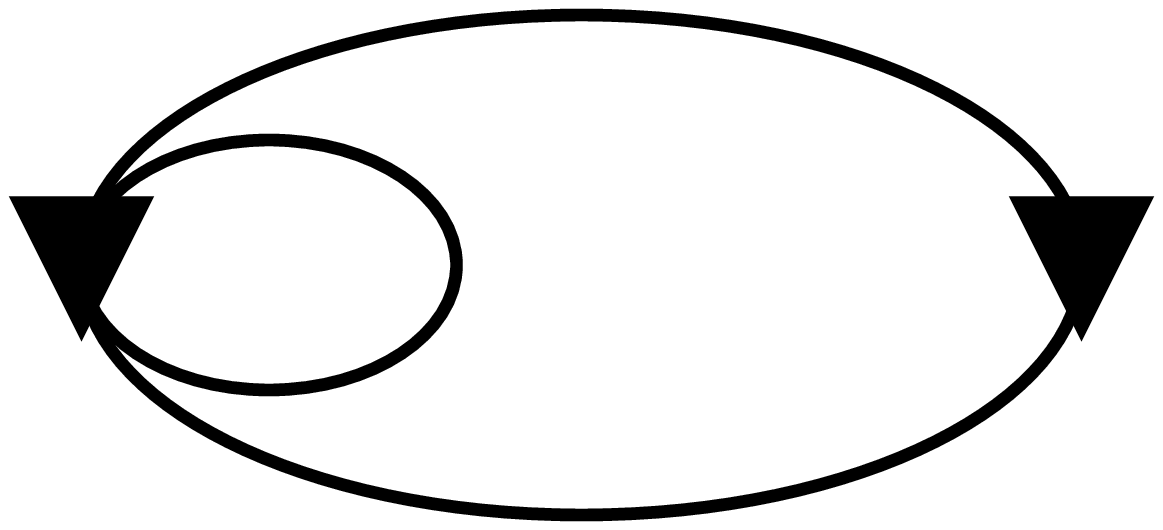} and \epsfxsize = 40pt \epsfysize = 20pt \epsffile{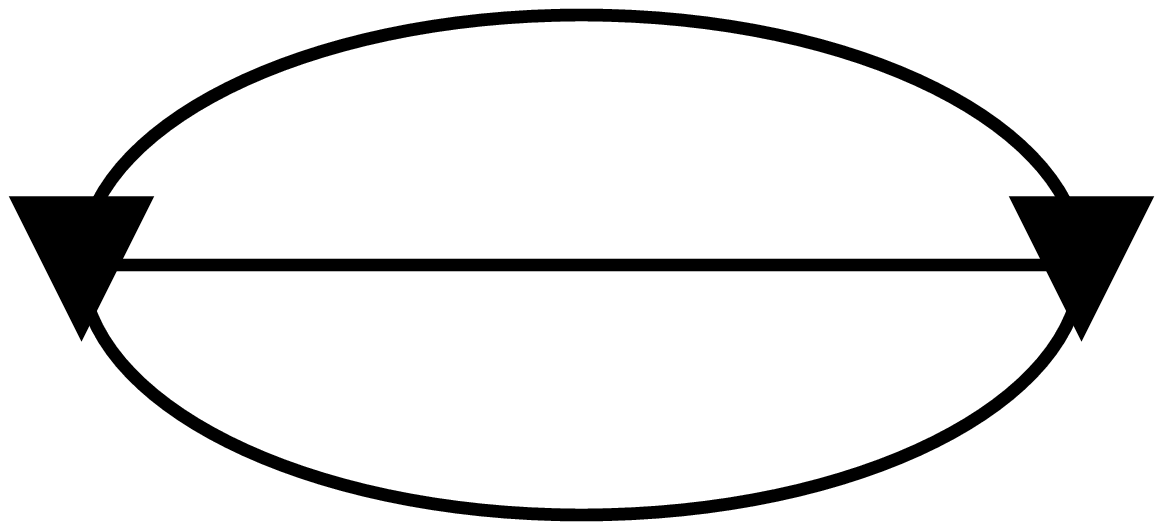} & 
&  & 16 &  & -4 & -2 \\ 
&  &  &  &  &  &  &  \\ \hline
&  &  &  &  &  &  &  \\ 
$\langle O_{2}^{(5)} S^{(3)}_{int} \rangle^{*} $ & \epsfxsize = 40pt %
\epsfysize = 20pt \epsffile{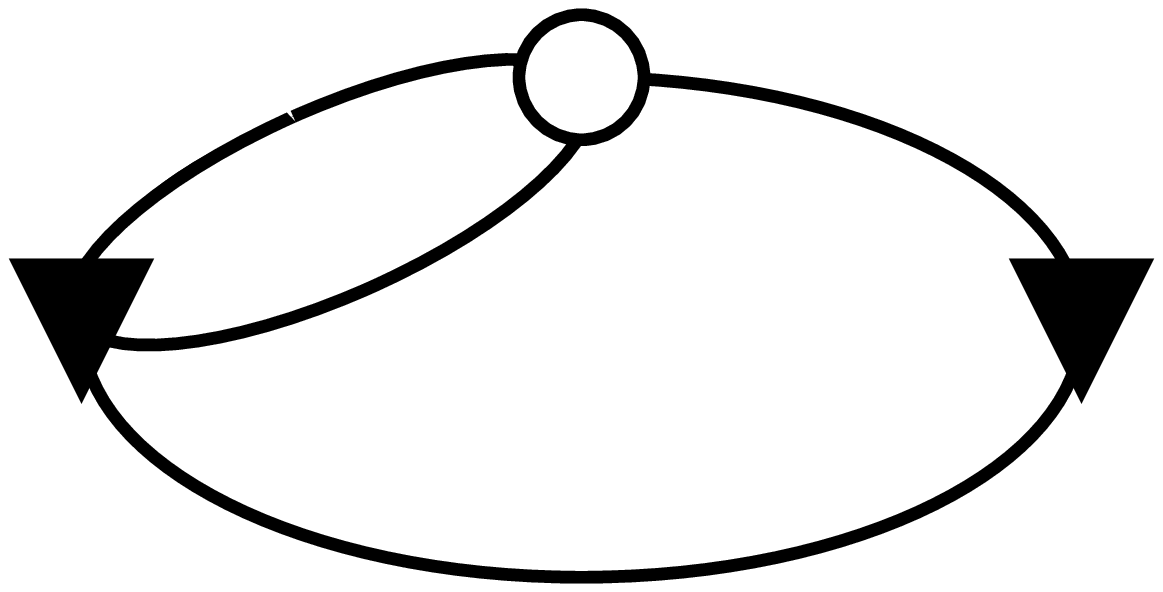} &  &  & -8 & 4 & 20 & 4 \\ 
&  &  &  &  &  &  &  \\ \hline
&  &  &  &  &  &  &  \\ 
$\langle O_{2}^{(4)} S^{(4)}_{0} \rangle $ & \epsfxsize = 40pt \epsfysize =
20pt \epsffile{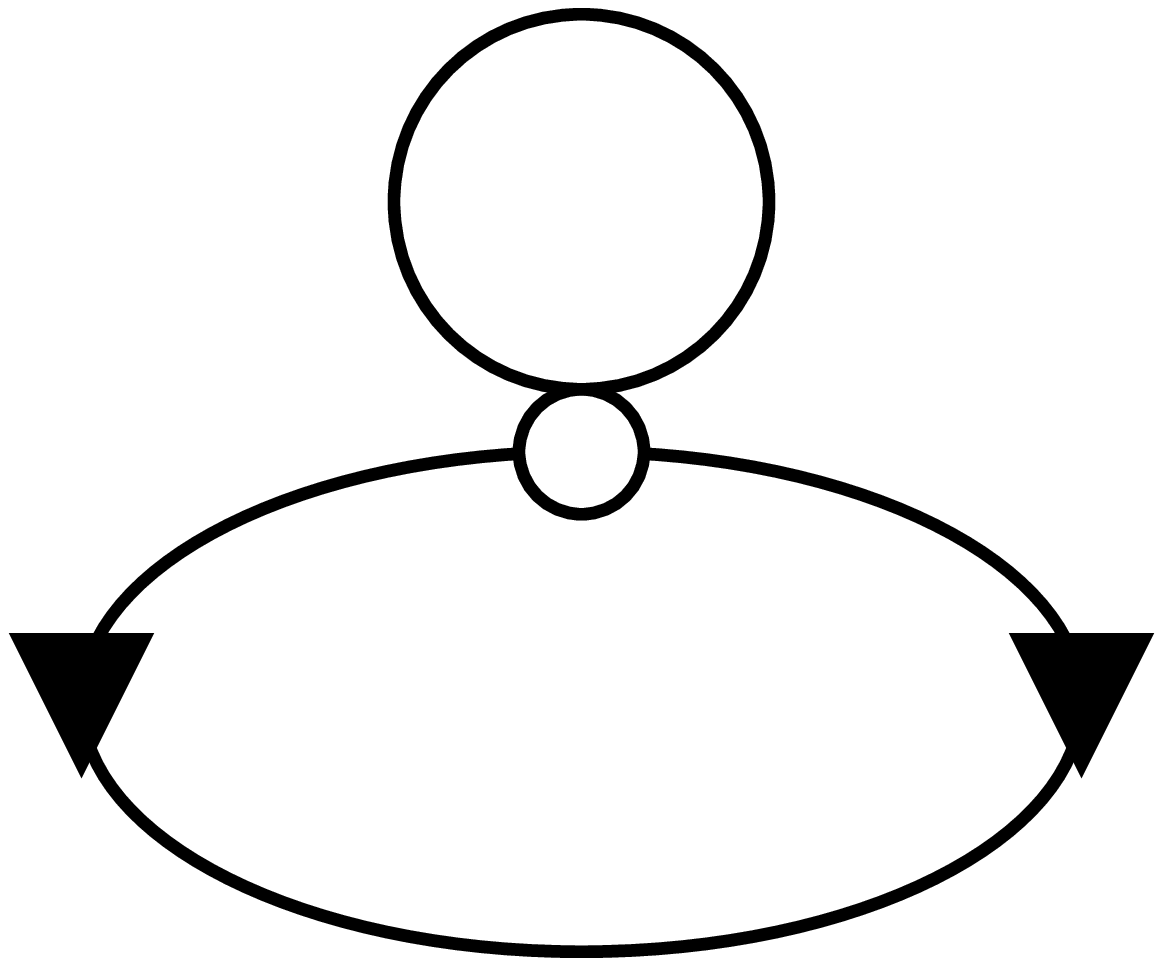} & 2 & 4 & -8 & -2 & 1 & 2 \\ 
&  &  &  &  &  &  &  \\ \hline
&  &  &  &  &  &  &  \\ 
$\langle O_{2}^{(2)}( S^{(4)}_{int} +\frac{1}{2} ( S^{(3)}_{int})^{2})
\rangle^{*} $ & \epsfxsize = 40pt \epsfysize = 20pt \epsffile{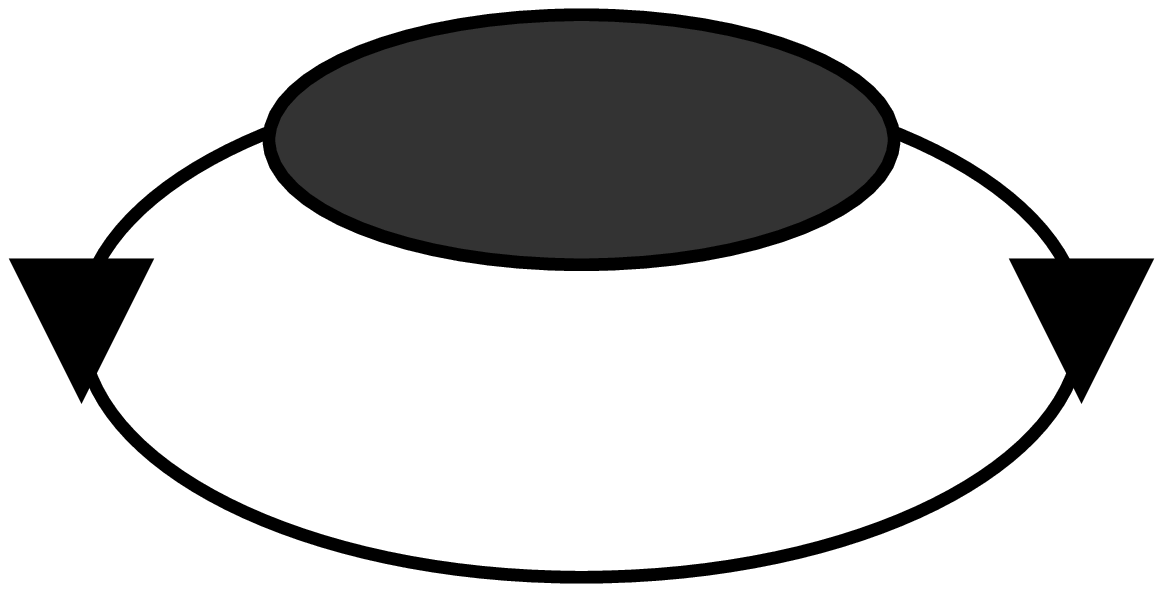} & -2
& -4 &  & -2 & -25 & -4 \\ 
&  &  &  &  &  &  &  \\ \hline\hline
&  &  &  &  &  &  &  \\ 
Total &  & 0 & 0 & 0 & 0 & - 8 & 0 \\ 
&  &  &  &  &  &  &  \\ \hline
\end{tabular}
\begin{figure}[tbp]
\end{figure}
\caption{ The second-loop contributions to the $O_{2} $ term in the
effective conductivity. The $\alpha$-dependent and $1/\epsilon^{2}$
contributions. The symbol $^{*}$ denotes that we exclude integrals $A_{0}$
and $C_{0}^{^{\prime}}$ which cancel in the sum of the two terms. A black
solid triangle denotes the current vertex in $O_{2} $ term, a white solid
dot denotes the vertex in $S $ terms, and }
\begin{figure}
\epsfxsize = 200pt
\epsfysize = 30pt
\epsffile{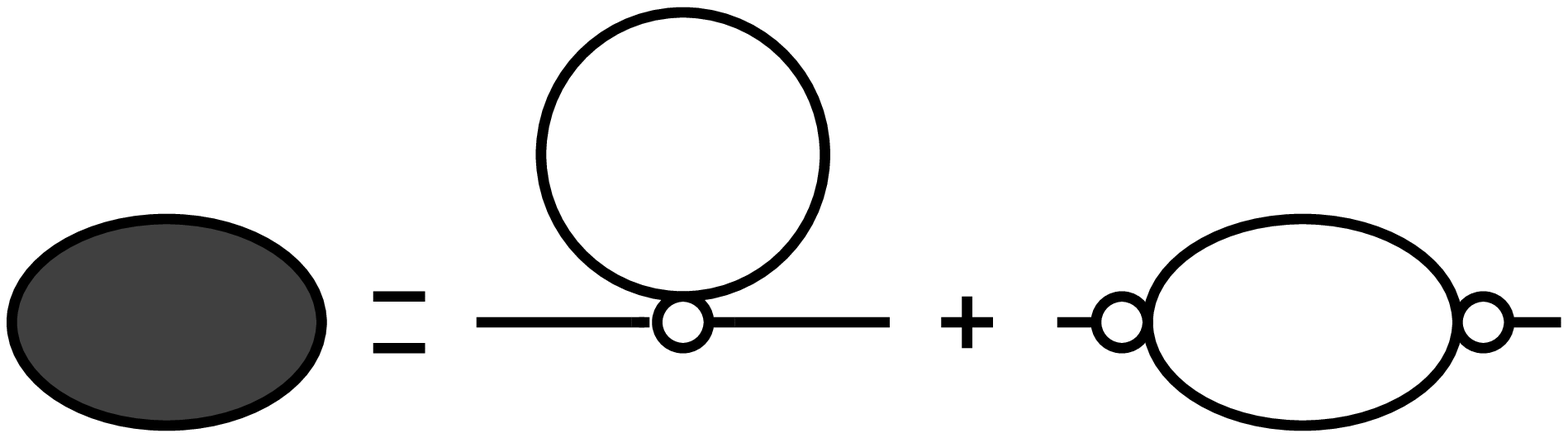}
\end{figure}
\end{table}


\begin{table}[tbp]
\begin{tabular}{|c|c||c|}
\hline
&  &  \\ 
Contractions & Diagrams & $\frac{1}{\epsilon } $ \\ 
&  &  \\ \hline\hline
&  &  \\ 
$\langle O^{(4)}_{1} \rangle $ & \epsfxsize = 40pt \epsfysize = 20pt %
\epsffile{o4.ps} & $0 $ \\ 
&  &  \\ \hline
&  &  \\ 
$\langle O_{1}^{(3)} S^{(3)}_{int} \rangle $ & \epsfxsize = 40pt \epsfysize %
= 20pt \epsffile{o3s3.ps} & $8 + 4 \zeta(3) $ \\ 
&  &  \\ \hline
&  &  \\ 
$\langle O_{1}^{(2)}( S^{(4)}_{int} + S^{(4)}_{0} +\frac{1}{2} (
S^{(3)}_{int})^{2}) \rangle $ & \epsfxsize = 40pt \epsfysize = 20pt %
\epsffile{o2s4.ps} & $- 4 - 2\zeta(3) - 2 \pi^{2} / 3 $ \\ 
&  &  \\ \hline\hline
&  &  \\ 
$\langle O^{(6)}_{2} \rangle $ & \epsfxsize = 40pt \epsfysize = 20pt %
\epsffile{o6.ps} & $- \frac{\pi^{2}}{3} + \frac{ \pi^{2}}{2} \log 2 + \frac{
\pi^{4}}{12} + \frac{11 \zeta(3)}{2} + \frac{\pi^{2}}{3} \log^{2} 2 - \frac{1%
}{3}\log^{4} 2 $ \\ 
& \epsfxsize = 40pt \epsfysize = 20pt \epsffile{oo6.ps} & $- 7 \zeta(3) \log
2 - 8 Li_{4}(\frac{1}{2}) $ \\ 
&  &  \\ \hline
&  &  \\ 
$\langle O_{2}^{(5)} S^{(3)}_{int} \rangle^{*} $ & \epsfxsize = 40pt %
\epsfysize = 20pt \epsffile{o5s3.ps} & $-12 + 4 \zeta(3) + 4 \pi^{2} / 3 $
\\ 
&  &  \\ \hline
&  &  \\ 
$\langle O_{2}^{(4)} S^{(4)}_{0} \rangle $ & \epsfxsize = 40pt \epsfysize =
20pt \epsffile{o4s4.ps} & $44 / 3 $ \\ 
&  &  \\ \hline
&  &  \\ 
$\langle O_{2}^{(2)}( S^{(4)}_{int} +\frac{1}{2} ( S^{(3)}_{int})^{2})
\rangle^{*} $ & \epsfxsize = 40pt \epsfysize = 20pt \epsffile{o4ss4.ps} & $%
55 /2 - 2 \zeta(3) - \frac{8}{3} \pi^{2} + 12 \log^{2} 2 - 44 \log 2 +16 G -
8 Li_{2}(\frac{1}{2}) $ \\ 
&  &  \\ \hline\hline
&  &  \\ 
Total &  & $34 + \frac{1}{6} - 3 \pi^{2} + \frac{19}{2}\zeta(3) +16 \log^{2}
- 44 \log 2 + \frac{\pi^{2} }{2} \log 2 +16 G $ \\ 
&  & $+ \frac{\pi^{4}}{12} + \frac{\pi^{2}}{3} \log^{2} 2 - \frac{1}{3}
\log^{4} 2 - 7 \zeta(3) \log 2 - 8 Li_{4}(\frac{1}{2}) $ \\ 
&  &  \\ \hline
\end{tabular}
\begin{figure}[tbp]
\end{figure}
\caption{ The second-loop contributions to the $O_{2} $ term in the
effective conductivity. The $1/\epsilon$ contributions. The symbol $^{*}$
denotes that we exclude integrals $A_{0}$ and $C_{0}^{^{\prime}}$ which
cancel in the sum of the two terms. A black solid dot denotes the vertex in $%
O_{1} $ term, a black solid triangle denotes the current vertex in $O_{2} $
term, a white solid dot denotes the vertex in $S $ terms, and }
\begin{figure}
\epsfxsize = 200pt
\epsfysize = 30pt
\epsffile{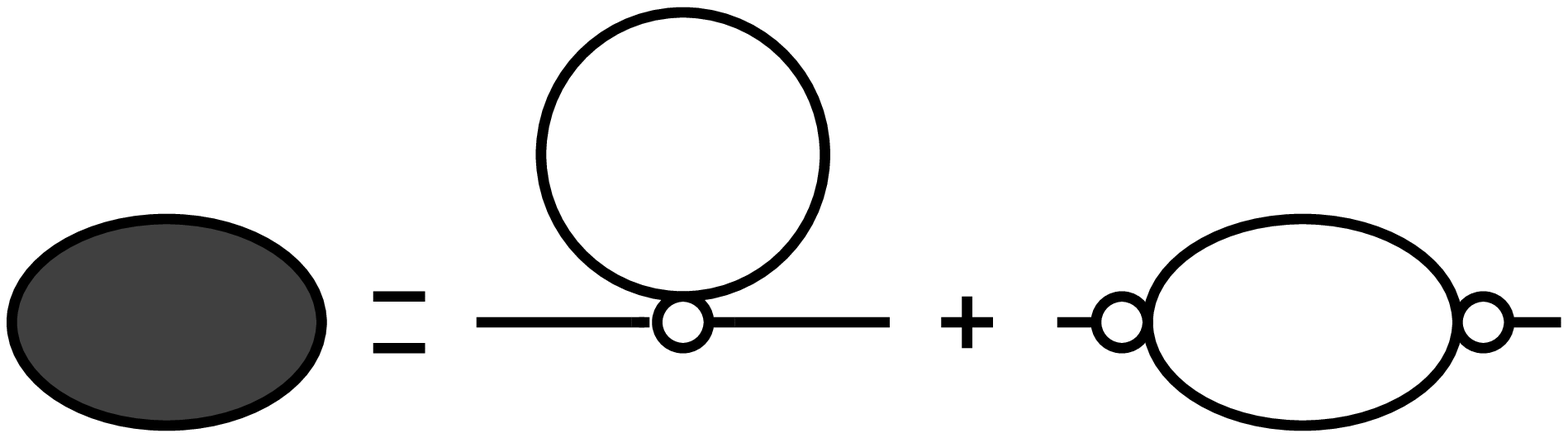}
\end{figure}
\end{table}

\begin{multicols}{2}

\section{Dynamical scaling}

\label{SR}

\subsection{Relation between $h^{^{\prime }}$ and $\omega _{s}$}

In the Section we combine the two loop computations of this paper with those
of the amplitude $z_{0}$ and establish the connection between the effective
mass $h^{^{\prime }}$ and the frequency $\omega _{s}$. For this purpose,
recall that the renormalization of the $z$ field was obtained from the
derivative of the free energy $F$ (or rather, the grand canonical potential)
with respect to $\ln T$.\cite{bps2} The result of the computation was as
follows 
\begin{equation}
\frac{dF}{d\ln T}=2\sum_{s>0}\omega _{s}z_{0}M_{b}(t_{0},h_{s}^{2}),
\label{derF}
\end{equation}
where 
\begin{equation}
M_{b}(t_{0},h_{s}^{2})=1+\frac{h_{s}^{2\epsilon }t_{0}}{2\epsilon }+\frac{%
h_{s}^{4\epsilon }t_{0}^{2}}{\epsilon ^{2}}\left( -\frac{1}{8}+\epsilon (%
\frac{1}{4}+\frac{\pi ^{2}}{24})\right) .  \label{M}
\end{equation}
Here, the frequency enters through the quantity $h_{s}^{2}=\kappa ^{2}z_{0}s=%
\frac{2\pi }{\Omega _{d}}\omega _{s}z_{0}t_{0}$ which has the dimension of
mass squared. The frequency dependence in $\sigma ^{^{\prime }}(s)$ is
restored by writing 
\begin{equation}
\sigma ^{^{\prime }}(s)=\frac{4\Omega _{d}}{t_{0}}R_{b}(t_{0},h_{s}^{2})
\label{R}
\end{equation}
with

\begin{equation}
R_{b}(t_{0},h_{s}^{2})=1+\frac{h_{s}^{2\epsilon }t_{0}}{\epsilon }+(A-1/2)%
\frac{h_{s}^{4\epsilon }t_{0}^{2}}{\epsilon }  \label{Rr}
\end{equation}
One can easily verify that Eqs. (4.1--4.4) lead to the same expressions for
$Z_{1}$ and $Z_{2}$ and, hence, the same $\beta $ and $\gamma $ functions as
those of the previous Section. Eq. (4.4) is therefore the correct result.

The relation between $h^{^{\prime}}$ and $\omega_s$ can now be made more
explicit by writing 
\begin{equation}
h^{^{\prime}2} = h_{s}^{2} M_b (t_{0},h_{s}^{2})/ R_b (t_{0},h_{s}^{2}).
\end{equation}
Here, $h^{^{\prime}}$ is the effective mass that is induced by the frequency 
$\omega_s$ and the result is consistent with all previous statements and
explicit computations.\cite{bps2}

\subsection{The Goldstone phase}

\label{MIT}

\subsubsection{Specific heat and AC conductivity}

The zero of the $\beta $ function, Eq. (\ref{pf}), determines a critical
point $t_{c}=O(\epsilon )$ that separates the Goldstone or metallic phase ($%
t<t_{c}$) from an insulating phase ($t>t_{c}$). To second order in $\epsilon 
$ we have 
\begin{equation}
t_{c}=\epsilon -2A\epsilon ^{2}\approx \epsilon -3.28\epsilon ^{2}
\end{equation}
We see that the $\epsilon ^{2}$ contribution is rather large and the
expansion can clearly not be trusted for $\epsilon =1/2$ or three spatial
dimensions. This is a well-known drawback of asymptotic expansions and the
two-loop theory is otherwise necessary to completely establish the scaling
behavior of the electron gas in $2+2\epsilon $ spatial dimensions. To
discuss this scaling behaviour, we proceed and express Eqs. (\ref{derF}) and
(\ref{R}) in terms of the renormalized parameters $t$ and $z$. The results
can be written in the following general form

\begin{eqnarray}
\frac{dF}{d\ln T} &=&2\sum_{s>0}\mu ^{2\epsilon }\omega _{s}zM(t,\omega
_{s}z),  \label{derFren} \\
\sigma ^{^{\prime }}(s) &=&\mu ^{2\epsilon }\frac{4\Omega _{d}}{t}R(t,\omega
_{s}z).  \label{sigmaren}
\end{eqnarray}
The expressions are now finite in $\epsilon $. The AC conductivity is
obtained from $\sigma ^{^{\prime }}(s)$ by replacing the imaginary
frequencies $i\omega _{s}$ by real ones $\omega $. On the other hand, the
specific heat of the electron gas can be expressed as \cite{bps2} 
\begin{equation}
c_{v}=\int_{0}^{\infty }d\omega \frac{\partial f_{BE}}{\partial T}\omega
\rho _{qp}(\omega ),  \label{cv}
\end{equation}
where 
\begin{equation}
f_{BE}=\frac{1}{e^{\omega /T}-1}
\end{equation}
and 
\begin{equation}
\rho _{qp}(\omega )=\frac{z}{\pi }(M(t,i\omega z)+M(t,-i\omega z))
\end{equation}
is the density of states of bosonic quasiparticles indicating that the
Coulomb system is unstable with respect to the formation of particle-hole
bound states. \cite{es}

\subsubsection{Scaling results}

Next, from the method of characteristics we can obtain the general scaling
behavior of the quantities $M$ and $R$ as usual: 
\begin{eqnarray}
M(t,\omega _{s}z) &=&M_{0}(t)G(\omega _{s}z\xi ^{d}M_{0}(t)),  \nonumber \\
R(t,\omega _{s}z) &=&R_{0}(t)H(\omega _{s}z\xi ^{d}R_{0}(t)).
\label{scale}
\end{eqnarray}
Here $G$ and $H$ are unspecified functions, whereas $\xi $, $R_{0}$ and $%
M_{0}$ each have a clear physical significance and are identified as the
correlation length, the DC conductivity and $\rho _{qp}(0)$ respectivily.
They obey the following equations 
\begin{eqnarray}
(\mu \partial _{\mu }+\beta \partial _{t})\xi (t) &=&0,  \nonumber \\
(\beta \partial _{t}-2\epsilon -\beta /t)R_{0}(t) &=&0,  \nonumber \\
(\beta \partial _{t}+\gamma )M_{0}(t) &=&0.
\end{eqnarray}
In the metallic phase ($t<t_{c}$) the solutions can be written as follows 
\begin{equation}
R_{0}(t)=(1-t/t_{c})^{2\epsilon \nu }\,\,,\,\,M_{0}(t)=(1-t/t_{c})^{\beta
_{0}},
\end{equation}
\begin{equation}
\xi =\mu ^{-1}t^{1/2\epsilon }(1-t/t_{c})^{-\nu },
\end{equation}
where the critical exponents $\nu $ and $\beta _{0}$ are obtained as 
\begin{equation}
\nu ^{-1}=\beta ^{^{\prime }}(t_{c})\,\,,\,\,\beta _{0}=-\nu \gamma (t_{c}).
\end{equation}
To second order in $\epsilon $ the results are 
\begin{eqnarray}
\nu ^{-1} &=&2\epsilon (1+2A\epsilon )\approx 2\epsilon +6.56\epsilon ^{2} 
\nonumber \\
\beta _{0} &=&\left( 1+(\pi ^{2}/6+3-4A)\epsilon \right) /2\approx
0.50-0.96\epsilon .
\end{eqnarray}
Both the DC conductivity $R_{0}$ and the quantity $M_{0}$ vanish as one
approaches the metal-insulator transition at $t_{c}$. The results are quite
familiar from the Heisenberg ferromagnet where $M_{0}$ stands for the
spontaneous magnetization. Unlike the free electron gas,\cite{inst}
however, the interacting system with Coulomb interactions has a true order
parameter, $M_{0}$, which is associated with a non-Fermi liquid behavior of
the specific heat.

\subsubsection{Equations of state}

The explicit results of Section A can be used to completely determine the
quantities $M$ and $R$ in the Goldstone phase. They take the form of an
''equation of state'' \cite{pw} 
\begin{equation}
\frac{\omega _{s}zt}{M^{\delta }}=\left( \frac{t_{c}}{t}\right) ^{1/\epsilon
}\left( 1+(2\epsilon \nu -1)(1-\frac{t}{t_{c}})-2\epsilon \nu \frac{1-t/t_{c}%
}{M^{1/\beta _{0}}}\right) ^{1/\epsilon },\nonumber
\end{equation}
\begin{equation}
\frac{\omega _{s}zt}{R^{\kappa }}=\left( \frac{t_{c}}{t}\right) ^{1/\epsilon
}\left( 1-\frac{1-t/t_{c}}{R^{1/2\epsilon \nu }}\right) ^{1/\epsilon }.
\label{eqs}
\end{equation}
Here, the exponents $\delta $ and $\kappa $ can be obtained from the values
of $\nu $ and $\beta _{0}$ following the relations 
\begin{equation}
d\nu =\beta _{0}(\delta +1)\,\,,\,\,2\epsilon \nu \kappa =\beta _{0}\delta .
\end{equation}
The universal features of the ''equations of state'' are the Goldstone
singularities at $t=0$ and the critical singularities near $t_{c}$. As for
the specific heat, we find the usual behavior $c_{v}=\gamma _{0}T$ at $t=0$
but at criticality the following algebraic behavior is found $c_{v}=\gamma
_{1}T^{1+1/\delta }$.

It is important to remark that the expression for the conductivity $R$ can
also be used in the case of finite temperatures and we may, on simple
dimensional grounds, substitute $T$ for $\omega _{s}$. The results, however,
strictly hold for the Goldstone and critical phases only. The ''equations of
state'' cannot be analytically continued and used to obtain
information on the insulating phase. As we already mentioned in the
introduction, the strong coupling phase is controlled by different operators
in the theory and has a distinctly different frequency and temperature
dependence.\cite{pb01}

\subsection{Plateau transitions in the quantum Hall regime}
\subsubsection{Introduction}
In this Section we briefly describe how the results of this paper are
extended to include the plateau transitions in the quantum Hall regime. For
this purpose recall that the theory in two spatial dimensions and strong
magnetic fields is given by

\begin{equation}
S[Q,A]\rightarrow S[Q,A]+\frac{\sigma _{xy}^{0}}{8}\int_{x}{\rm tr}\epsilon
_{ij}Q[D_{i},Q][D_{j},Q].
\end{equation}
The theory depends on the $\theta $ term, or $\sigma _{xy}$ term, in a
non-perturbative manner and the general form of the renormalization group
equations can now be written as \cite{eurolett} 
\begin{eqnarray}
\frac{d\sigma _{xx}}{d\ln \mu } &=&\beta _{xx}(\sigma _{xx},\sigma _{xy}), 
\nonumber \\
\frac{d\sigma _{xy}}{d\ln \mu } &=&\beta _{xx}(\sigma _{xx},\sigma _{xy}),
\label{RG} \\
\frac{d\ln z}{d\ln \mu } &=&\gamma (\sigma _{xx},\sigma _{xy}).  \nonumber
\end{eqnarray}

The interesting physics actually occurs in the strong coupling phase ($%
\sigma _{xx}<1$) where the crossover takes place from the perturbative regime of 
quantum interference effects, as studied in this paper, 
to the quantum Hall regime that generally appears in the limit of much larger 
distances only (Fig. 1).

\begin{figure}[tbp]
\epsfxsize=200pt
\epsfysize=200pt
\epsffile{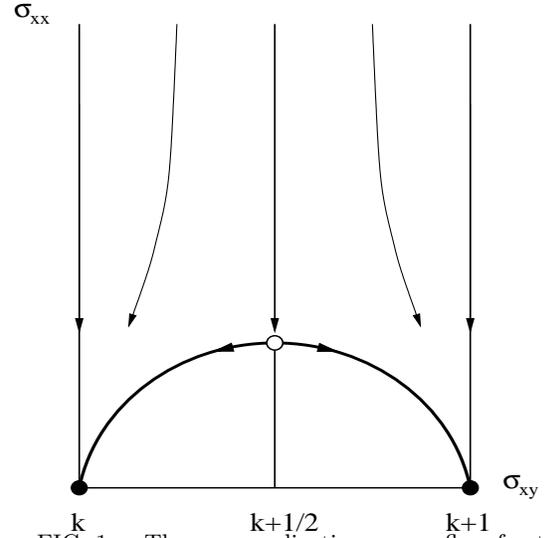}
\caption{
The renormalization group flow 
for the conductances. 
The arrows indicate 
the scaling towards the infrared.}
\end{figure}

As an important general remark we can say that the quantum Hall effect is a
universal, strong coupling feature of the $\theta $ term, or {\em instanton vacuum}, and
fundamental aspects of the problem have not been recognized until recently. We
mention in particular the fact that the theory displays {\em %
massless excitations} that always exist at the edge of the system. \cite{pbv}
This new ingredient turns out to have fundamental consequences for 
longstanding problems such as the {\em quantization of topological charge}, the
general meaning of {\em instantons} etc. Moreover, the concept of massless
chiral edge modes can be used to unravel some of the outstanding strong
coupling aspects of the theory such as the {\em exact quantization%
} of the Hall conductance which is represented by the 
infrared stable fixed points at $\sigma _{xx}=0$ and $\sigma _{xy}=k$
in the scaling diagram of Fig. 1. 

Perhaps more surprizingly, a gapless phase seems to always exist in the theory
at $\theta =\pi $ or $\sigma _{xy}^{0}$ equal to an half-integer. This
fundamental aspect of the quantum Hall effect is displayed even by the $%
CP^{N-1}$ theory with large values of $N$. \cite{pbv} These results indicate 
that the quantum Hall effect is a generic feature of the $\theta$ term in
asymptotically free field theory and, contrary to the previous believes,
the number of field components plays a secondary role only. Recall that
the free electron theory, the Finkelstein approach and the $CP^{N-1}$
model with large $N$ are all topologically equivalent. They have important
features in common such as asymptotic freedom and instantons. They are
only different in the manner the number of field components in the theory
is being handled. This does not affect the fundamentals of the
quantum Hall effect, however, but only the critical singularities at $\theta = \pi$
which are different in each case.

\subsubsection{Scaling of conductances}
We next focus on the consequences of the unstable fixed points in Fig. 1,
located at $\sigma _{xy}=k+\frac{1}{2}$ and $\sigma _{xx}=\sigma _{xx}^{*}$
which is of order unity. These fixed points describe the critical
singularities of the quantum Hall plateau transitions. \cite{eurolett}
A finite value of $\sigma _{xx}^{*}$ indicates that we are dealing with
a critical metallic state
which is much the same phenomenon as the metal-insulator transition that
separates the Goldstone phase from the insulating phase in the theory in $%
2+2\epsilon $ dimensions. The quantum Hall regime therefore provides a
unique laboratory inwhich the properties of disorder driven quantum phase 
transition can be explored and investgated in detail.


Let us first recall the results for the conductances 
$\sigma_{xx}^{^{\prime}}$ and 
$\sigma_{xy}^{^{\prime }}$ as obtained in Ref. 8

\begin{eqnarray}
\sigma _{xx}^{^{\prime }} &=&f_{xx}[(zT)^{-\kappa }(\nu _{B}-\nu _{B}^{*})], 
\nonumber \\
\sigma _{xy}^{^{\prime }} &=&f_{xy}[(zT)^{-\kappa }(\nu _{B}-\nu _{B}^{*})].
\end{eqnarray}
Here, the functions $f_{xx}(X)$ and $f_{xy}(X)$ are
regular (differentiable) functions for small $X$, $\nu_{B}=\sigma _{xy}^{0}\propto 1/B$
is the filling fraction of the Landau
levels and $\nu _{B}^{*}=k+1/2$ is the critical value, corresponding to the
center of the Landau band. The exponent $\kappa =p/2\nu \approx 0.42$ 
has been extracted
from the experimental transport data taken from low mobility
heterostructures in the quantum Hall regime. \cite{experiments}

Notice that the scaling variable $X$ can be expressed as 
$(h^{^{\prime }}\xi)^{-1/\nu}$ where $h^{^{\prime }}$ is the mass
that is induced by finite temperatures (or frequency) 
\begin{equation}
h^{^{\prime }}=(zT)^{p/2}\;,\;\;(z\omega)^{p/2}.
\end{equation}
The $\xi $ is the diverging correlation length at the center of the Landau
band 
\begin{equation}
\xi \propto |\nu _{B}-\nu _{B}^{*}|^{-\nu }=|\sigma _{xy}-k-\frac{1}{2}|^{-\nu}.
\label{xi}
\end{equation}
The critical exponent $\nu $ has the same meaning as before whereas $p$ was
originally introduced as the {\em inelastic scattering time} exponent.
\cite{freepart} Both
are defined formally by the $\beta _{xy}$ and $\gamma $ functions according to

\begin{eqnarray}
{\nu ^{-1}} &=&\partial \beta _{xy}^{*}/\partial \sigma _{xy},  \nonumber \\
p &=&1+\frac{1}{\delta }=\frac{1}{1+\gamma ^{*}/2},
\end{eqnarray}
where $\beta _{xy}^{*}=\beta _{xy}(\sigma _{xx}^{*},k+1/2)$ and $\gamma
^{*}=\gamma (\sigma _{xx}^{*},k+1/2)$.

\subsubsection{Particle-hole symmetry, duality}

Generally speaking, one expects
the functions $f_{xx} (X)$ and $f_{xy} (X)$ to be universal scaling functions,
describing the points on the renormalization group trajectory that connects 
the unstable fixed points with the stable ones (Fig. 1). \cite{freepart}
There is, however, interesting physics associated with this statement
of universality and the subject is an extremely important objective 
for experimental research.

The problem with the plateau transitions is that although the {\em macroscopic} 
conductances $\sigma_{xx}^{\prime}$ and $\sigma_{xy}^{\prime}$
are well defined and sharply distributed at finite $T$, this
is not the case for the {\em mesoscopic} conductances 
which are defined for finite lengthscales, 
of the order of the phase breaking length $1/h'$. 
The mesoscopic
conductances are, in fact, broadly distributed and the size of the fluctuations is
comparable or larger than the mean value. Since the $1/h'$ is the only length 
scale in the problem with Coulomb interactions and at finite $T$,
it directly follows that the relation between the {\em mesoscopic conductance
distributions} and the {\em measured} or {\em macroscopic conductance} 
must be non-trivial in general. 
For example, it is necessary to construct {\em block models} \cite{cohen}
that describe the electron transport process in terms of a classical network
of (mesoscopic) conductances that are randomly distributed over the different
areas ({\em blocks}) in the system of size $1/h'$. 

The concept of {\em block models} complicates such aspects like 
{\em particle-hole symmetry} that is displayed by the physical observables of the
electron gas. {\em Particle-hole symmetry}, just like the
quantization of the Hall conductance, is a direct consequence
of one of the most fundamental aspects of the instanton vacuum, 
namely {\em quantization} of {\em topological charge}. 
It can be expressed as follows
\begin{eqnarray}
f_{xx} (X) & = & f_{xx} (-X) \nonumber \\
f_{xy} (X) & = & 2k +1 - f_{xy} (-X)
\end{eqnarray}
More generally, one can show that {\em particle-hole symmetry}
is displayed by the entire distribution functions of the mesoscopic
conductances, rather than by the macroscopic quantities or averaged quantities alone.

It is clear that the theory of block models is particularly sensitive with
regard to the many controversial issues that presently span the subject of
mesoscopic fluctuations.\cite{altshuler} It is important to keep in mind
that the quantum Hall plateau transitions take place in precisely the regime 
($\sigma_{xx} < 1$) where not only the conductance fluctuation are 
uncontrolled, but also the infinite set of higher dimensional operators 
that enters in the definition of the higher order momenta of the distribution 
functions. Obviously, for the more difficult problems like quantum
criticality in the presence of the Coulomb interactions, one can not
{\em just assume} that the theory automatically takes care of itself in each
and every fronts.

Following Kivelson et al, \cite{kivel} however, one can proceed in a pragmatic
fashion and employ the Chern Simons mapping of abelian
quantum Hall states to show that the system has a {\em dual} symmetry.
Provided one works at finite $T$ and with system sizes that are much larger than
$1/h'$, the mapping of conductances is not affected by the
fluctuations that occur at mesoscopic lengthscales.\cite{pb01} 
By making furthermore use
of {\em particle-hole symmetry} and by identifying
the functions $f_{xx} (X)$ and $f_{xy} (X)$ as the subspace of conductances 
that is {\em dual} under the Chern Simons mapping, one arrives at the following
result
\begin{eqnarray}
f_{xx} (X) & = & \frac{g(X)}{1+g^2 (X)} \nonumber \\
f_{xy} (X) & = & k + \frac{1}{1+g^2 (X)}
\label{ff}
\end{eqnarray}
where the function $g(X) = e^{a_1 X + a_3 X^3 + ...}$ obeys the general contraint
\begin{equation}
g(X) =g^{-1} (-X)
\label{gg}
\end{equation}
These results imply that the sequence of plateau transitions in the quantum 
Hall regime ends up at $k=0$ in a so-called {\em quantum Hall insulating
phase} which means that the Hall resistance $\rho_{xy}$ remains quantized
throughout the lowest Landau level. 

It is important to remark that the statement of duality has been carried
out in a manner which is consistent with the gradient expansion that generally
defines the effective action or sigma model approach. \cite{bps1} If, on the other hand,
the effective action procedure were to fail and, say, terms of higher dimension
would generally become important, \cite{altshuler} then the statements made by
Eqs. (\ref{ff}) and (\ref{gg}) would clearly have no meaning and the theory of quantum transport 
must be largely reconsidered. 

With regard to the universality of the functions $f_{xx} (X)$ and $f_{xy} (X)$, the 
experimental situation has remained unresolved for a long time. 
However, recent
experiments have clearly demonstrated that Eqs. (\ref{ff}) and (\ref{gg}) are valid, at least
for the lowest Landau level. The transport data were taken 
from a low mobility $InGaAs/InP$ heterostructure in strong magnetic fields and 
at low temperatures.\cite{newexpt} 

The new results indicate that
the lack of universality, that was previously found,\cite{experiments1,experiments} 
is merely the consequence
of sample inhomogeneities. This means that there is little room left for 
the type of 
complications that arose in the perturbative theory of mesoscopic fluctuations. 
The experiments are in favor of {\em duality} as a fundamental symmetry of
the electron gas with Coulomb interactions. As shown by Eqs. (\ref{ff}) and (\ref{gg}),
this symmetry provides
fundamental support for the results of the renormalization theory.

It should be mentioned that the Chern Simons mapping of conductances can be 
carried out for almost any type of disorder and duality by itself
does therefore not provide
any garantee that the system is actually in a quantum critical state. For example, 
it is well known that complications arise in systems with longranged potential 
fluctuations and the matter has been extensive addressed in Ref. 11.
\subsubsection{Specific heat}
As we have mentioned earlier, it is necessary to identify other physical observables
in the problem
that can in principle be measured and used to extract the value of $p$ and $\nu$
separately. The microscopic theory of the electron gas in
$2+2\epsilon $ dimensions tells us that the natural quantity to consider is
the specific heat, Eq. (\ref{cv}). 
Moreover, we have shown in Ref. 8
that this quantity is unchanged under the Chern Simons mapping.

By using our general knowledge on the
renormalization group functions $\beta _{xx}$, $\beta _{xy}$ and $\gamma $
one can derive, in the
standard manner, the scaling form of the quantity $M(\sigma _{xx},\sigma
_{xy},\omega _{s}z)$ in the quantum Hall regime. This leads to the same
expression as in Eq. (\ref{scale}) with $M_0 (t)$ now replaced by 
$M_0 (\nu_B ) = |\nu_B -\nu_B^* |^{\beta_0}$ and $\xi$ given as in Eq. (25).
At the quantum
critical point ($\nu_B =\nu_B^*$)
we obtain the same non-Fermi
liquid expression as before

\begin{equation}
c_{v}=\gamma _{1}T^{p}.
\end{equation}
In different words, the physical observable, associated with the ''inelastic
scattering'' exponent $p$ in quantum Hall systems, is none other than the 
{\em specific heat} of the electron gas. A measurement of $c_{v}$ should
therefore provide the ultimate test on the consistency of the theory. 
This information is not present as of yet.

\section{ Conclusion}

\label{Conc}

In this paper we have completed the two-loop analysis of the Finkelstein
theory with the singlet interaction term. We have reported the detailed
computations of the conductivity which is technically the most difficult
part of the analysis. We have benifitted from the regularization procedure
involving the $h_0$ field, which has substantially simplified the two-loop
computations. 
Moreover, we have
obtained a general relation between the effective masses that are being
induced by the $h_0$ field on the one hand, and the frequency $\omega_n$ on
the other. This enables one to re-express the final answer in terms of
finite frequencies and/or temperature, simply by a substitution of the $h_0$
regulating field.

By combining the concept of ${\cal F}$ invariance with technique of
dimensional regularization, we have extracted new physical information on
the disordered electron gas with Coulomb interactions in low dimensions. In
particular, we now have a non-Fermi liquid theory for the specific heat and
dynamical scaling. 

The metal-insulator transition in $2+2\epsilon $
dimensions sets the stage for the plateau transitions in the quantum Hall
regime. We have identified the specific heat $c_{v}$ as the physical
observable that determines the exponent $p$, previously introduced as the
exponent for "inelastic scattering."

As a final remark we can say that our knowledge of the theory is limited
only by the accuracy with which one can give a numerical estimate of the
critical exponents $\nu $ and $p$. Except for the fact that $p$ is bounded
by $1<p<2$, \cite{pb01} the detailed values of $\nu $ and $p$ can only
be obtained by performing the renormalization group numerically. Notice that
the situation is somewhat similar for the metal-insulator transition in 
$2+2\epsilon $ dimensions. In that case, the limitations of the $\epsilon 
$expansion prevent us from having accurate exponents for the electron gas in
three spatial dimensions.

\section{Acknowledgement}

\label{Ack} We are indebted to E. Br\'{e}zin and A. Finkelstein for numerous 
conversations. One of us (I.B.)
is grateful to M. Feigel'man, M. Lashkevich, D. Podolsky 
and P. Ostrovsky for stimulating discussions. 
The research has been supported in
part by the Dutch Science Foundation FOM and by INTAS (Grant 99-1070). 

\setcounter{equation}{0}
\renewcommand{\theequation}{A.\arabic{equation}}
\section{Appendix A}

\label{AA} In this Appendix we present the final results for the various
integrals listed in Eqs. (\ref{start})-(\ref{end}). We shall follow the same
methodology as used in the two-loop computation of Ref. 10 and
employ the standard representation for the momentum and frequency integrals
in terms of the Feynman variables $x_{1}$, $x_{2}$ and $x_{3}$. We classify
the different contributions in Eqs. (\ref{start})-(\ref{end}) in different
catagories, labeled $A$-integrals, $B$-integrals etc. In total we have seven
different catagories, i.e. $A$, $B$, $C$, $D$, $H$, $S$ and $T$
respectively, which are discussed separately in Sections $A$ - $G$ of this
Appendix. The last Section, $H$, contains a list of abbreviations and a list
of symbols for those integrals that need not be computed explicitly because
their various contributions sum up to zero in the final answer.

In Appendix $B$ we present the main computational steps for a specific
example, the so-called $A_{10}$-integral. We show how the integral
representation of hypergeometric functions can be used to define both the $%
\epsilon $ expansion and the limit where $\alpha \rightarrow 0$.

\subsection{ The A - integrals}

\label{A}

\subsubsection{Definition}

To set the notation, we consider the integral

\begin{eqnarray}
X_{\nu,\eta}^\nu & = & -\frac{2^{1+\nu} a^{2+\mu} }{\sigma_{0} d^{\nu}} \int
\limits_{p q} p^{2 \nu} \sum_{k,m>0} m^{\mu}  \nonumber \\
& & D_{p+q}^{c}(m) D D^{c}_{p}(k) D^{1+\mu+\eta}_{q}(k+m) .  \label{pr}
\end{eqnarray}

Here, the three indices $\mu$, $\nu$ and $\eta$ generally take on the values 
$0, 1$. We shall only need those quantities $X_{\nu,\eta}^\nu$ which have $%
\eta = \nu$, however.

Using the Feynman trick, one can write (for the notation, see Section $H$)

\begin{eqnarray}
X_{\nu,\eta}^\nu & = & -\frac{2^{1+\nu} a^{2+\mu} }{\sigma_{0} d^{\nu}} \int
\limits_{p q} p^{2} \int \limits_{0}^{\infty} dm \; m^{\mu} \int
\limits_{0}^{\infty} dk  \nonumber \\
& & \frac{\Gamma(\mu +\eta+4)}{\Gamma(\mu + \eta+1)} \int
\limits_{\alpha}^{1} dz \int [] \; x_{2} x_{3}^{\mu +\eta}  \nonumber \\
& & \left[ \right . h_{0}^{2} + q^{2} x_{12} + p^{2} x_{13} + 2 {\bf p}
\cdot {\bf q} \; x_{1}  \nonumber \\
& & + a m ( \alpha x_{1} + x_{3}) + a k (z x_{2} + x_{3} ) \left . \right
]^{-\mu - \eta - 4}
\end{eqnarray}
Next, by shifting $q \rightarrow q - p x_{1} / x_{12}$, we can decouple the
vector variables ${\bf p}$ and ${\bf q}$ in the denominator. The integration
over $k, m, p$ and $q$ then leads to an expression that only involves the
integral over $z$ and the Feynman variables $x_1$, $x_2$ and $x_3$. Write

\begin{eqnarray}
X_{\nu,\eta}^\nu = \frac{ \Omega_{d}^{2} h_{0}^{4 \epsilon}}{ \sigma_{0}
\epsilon} A^{\nu}_{\mu,\eta}
\end{eqnarray}

then 
\[
A_{\mu \eta }^{\nu }=\int\limits_{\alpha }^{1}dz\int []\frac{%
x_{2}x_{3}^{1+\mu +\eta }(x_{1}+x_{2})^{\nu }(x_{i}x_{j})^{-1-\nu -\epsilon }%
}{(zx_{2}+x_{3})(\alpha x_{1}+x_{3})^{1+\mu }}. 
\]

To complete the list of $A$-integrals, we next define quantities that carry
either two indices $\mu , \nu$ or only a single index $\mu$. Like $A_{\mu
\eta}^{\nu}$, they all describe contractions that contain both momentum and
frequency integrals. The results are all expressed in terms of integrals
over $z$, $x_1$, $x_2$ and $x_3$.

\begin{eqnarray}
A_{\nu \mu } &=&\int\limits_{\alpha }^{1}dz(z-\alpha )^{1+\nu -\mu } 
\nonumber \\
&\times &\int []\frac{x_{1}^{\mu }x_{2}^{2+\nu -\mu }x_{3}^{\mu
}(x_{1}+x_{3})^{1-\mu }(x_{i}x_{j})^{-2-\epsilon }}{(\alpha
x_{1}+x_{3})^{1+\nu }(zx_{2}+x_{3})},
\end{eqnarray}
\begin{equation}
A_{0}=\int\limits_{\alpha }^{1}dz(z-\alpha )\int []\frac{%
x_{2}^{2}x_{1}(x_{i}x_{j})^{-2-\epsilon }}{(x_{3}+zx_{2})(zx_{2}+\alpha
x_{1}+2x_{3})},
\end{equation}
\begin{eqnarray}
A_{1} &=&\int\limits_{\alpha }^{1}dz(z-\alpha )^{2}\int []\frac{%
x_{2}^{3}(x_{1}+x_{3})(x_{2}+x_{3})}{(zx_{2}+x_{3})^{2}}  \nonumber \\
\times ( &x_{i}&x_{j})^{-2-\epsilon }\left( {}\right. \frac{1}{(\alpha
x_{1}+x_{3})^{2}}-\frac{1}{(zx_{2}+\alpha x_{1}+2x_{3})^{2}}\left. {}\right)
,
\end{eqnarray}
\begin{eqnarray}
A_{2} &=&\int\limits_{\alpha }^{1}dz(z-\alpha )(1-z)  \nonumber \\
&\times &\int []\frac{x_{2}^{3}(x_{1}+x_{3})(x_{i}x_{j})^{-2-\epsilon }}{%
(zx_{2}+x_{3})(\alpha x_{1}+x_{3})(zx_{2}+\alpha x_{1}+2x_{3})}
\end{eqnarray}
\[
A_{3}=\int\limits_{\alpha }^{1}dz(z-\alpha )\int []\frac{%
x_{2}^{2}(x_{1}+x_{3})(x_{i}x_{j})^{-2-\epsilon }}{(\alpha
x_{1}+zx_{2}+2x_{3})(zx_{2}+x_{3})}. 
\]

\subsubsection{$\epsilon$ expansion}

The calculation of integrals is staightforward but tedious and lengthy. Here
we only present the final results of those quantities that are needed. The
list does not contain the final answer for the $A_0$-integral because the
various contributions to $A_0$ sum up to zero in the final answer. These
same holds for some other integrals that are defined in Section $H$ and that
we do not specify any further.

\begin{eqnarray}
A_{00}^{0} &=&-\frac{\ln ^{2}\alpha }{\epsilon }+\zeta (3),  \nonumber \\
A_{10}^{0} &=&-\frac{\ln ^{2}\alpha +\ln \alpha }{\epsilon }-\frac{\ln
^{2}\alpha }{2}+\frac{\pi ^{2}}{6}+\zeta (3),  \nonumber \\
A_{01}^{1} &=&\frac{\ln \alpha }{\epsilon }-\frac{\ln ^{2}\alpha }{2}-2\ln
\alpha -\frac{\pi ^{2}}{3}+1,  \nonumber \\
A_{11}^{1} &=&\frac{\ln \alpha }{\epsilon }-\frac{\ln ^{2}\alpha }{2}-2\ln
\alpha -\frac{\pi ^{2}}{3},  \nonumber \\
A_{00} &=&\frac{\ln \alpha }{\epsilon }+\frac{\ln ^{2}\alpha }{2}+2\ln
\alpha +\frac{\pi ^{2}}{3}-1,  \nonumber \\
A_{10} &=&-\frac{1}{\alpha }-\frac{2\ln \alpha +3}{\epsilon }-\ln ^{2}\alpha
-5\ln \alpha -\frac{2\pi ^{2}}{3}+3,  \nonumber \\
A_{01} &=&-\ln \alpha -\frac{\pi ^{2}}{6}+1,  \nonumber \\
A_{11} &=&\frac{\ln \alpha +2}{\epsilon }+\frac{\ln ^{2}\alpha }{2}+3\ln
\alpha +\frac{\pi ^{2}}{2},
\end{eqnarray}
\begin{eqnarray}
A_{1} &=&-\frac{2}{\alpha }+\frac{2\ln ^{2}\alpha +4\ln \alpha }{\epsilon }%
-3\ln ^{2}\alpha  \nonumber \\
&+&8\ln 2\ln \alpha -\frac{17}{2}\ln \alpha +4K_{1}(\alpha
)+8J_{3}^{^{\prime }}(\alpha )  \nonumber \\
&-&\pi ^{2}-2\zeta (3)-6\ln ^{2}2+10\ln 2-\frac{1}{2},
\end{eqnarray}
\begin{eqnarray}
A_{2} &=&-\frac{\ln ^{2}\alpha +2\ln \alpha }{\epsilon }-2\ln \alpha -3\ln
2\ln \alpha  \nonumber \\
&-&J_{1}(\alpha )-K_{1}(\alpha )-2J_{3}^{^{\prime }}(\alpha )+A_{0}-\frac{%
\pi ^{2}}{6}  \nonumber \\
&+&1+\zeta (3)+3\ln ^{2}2-3\ln 2-3Li_{2}(1/2),
\end{eqnarray}
\begin{equation}
A_{3}=A_{0}-2Li_{2}(\frac{1}{2})+\frac{\pi ^{2}}{6}.
\end{equation}

\subsection{ The B - integrals}

\label{B}

\subsubsection{Definition}

The $B$-integrals are similarly defined in terms of the variables $z$, $%
x_{1} $, $x_{2}$ and $x_{3}$. However, they describe only those contractions
that contain frequency sums and no momentum integrals. 
\begin{equation}
B_{\mu }=\int\limits_{\alpha }^{1}\frac{dz}{z^{\mu }}\int []\frac{x_{1}^{\mu
-1}x_{2}x_{3}^{-\mu -\epsilon }(x_{1}+x_{2})^{-\mu -\epsilon }}{(\alpha
x_{2}+zx_{3}+x_{1})},
\end{equation}

\subsubsection{$\epsilon$ expansion}

\begin{eqnarray}
B_{1} &=&\frac{\ln \alpha }{\epsilon }+\frac{\ln ^{2}\alpha }{2}+\ln \alpha ,
\nonumber \\
B_{2} &=&-\frac{1}{\alpha }+\frac{\ln ^{2}\alpha }{\epsilon }+\frac{2\ln
\alpha }{\epsilon }-2\ln \alpha -2.
\end{eqnarray}

\subsection{The C - integrals}

\label{C}

\subsubsection{Definition}

The $C$-integrals contain one additional integration over $y$, besides the
ones over $z$ and the Feynman variables $x_1$, $x_2$ and $x_3$. They
originate from expressions involving integrations over both frequencies and
momenta.

We distingish between quantities with two indices $\mu $ and $\nu $ 
\begin{equation}
C_{\mu \nu }=\int\limits_{\alpha }^{1}dzdy\int []\frac{x_{1}^{\mu
}x_{2}x_{3}(x_{2}+x_{3})^{1-\mu }(x_{i}x_{j})^{-2-\epsilon }}{%
(zx_{3}+x_{1})(yx_{2}+x_{1})^{\nu }(zx_{3}+yx_{2})^{1-\nu }}
\end{equation}
and those that carry only a single index $\nu $ 
\begin{eqnarray}
C_{\nu } &=&\int\limits_{\alpha }^{1}dz(1-z)^{\nu }\int\limits_{\alpha
}^{1}dy  \nonumber \\
&\times &\int []\frac{x_{2}^{2-\nu }x_{3}^{1+\nu }(x_{1}+x_{2})^{\nu
}(x_{i}x_{j})^{-2-\epsilon }}{(zx_{3}+x_{1})(yx_{2}+x_{1})^{\nu
}(zx_{3}+yx_{2}+2x_{1})}.
\end{eqnarray}

\subsubsection{$\epsilon$ expansion}

\begin{eqnarray}
C_{00} &=&\frac{\ln \alpha }{\epsilon }-\frac{\ln ^{2}\alpha }{2}-2\ln
\alpha +\frac{\pi ^{2}}{4}\ln 2-\frac{\pi ^{2}}{6}+\frac{15}{4}\zeta (3) 
\nonumber \\
&-&\frac{\pi ^{4}}{24}-\frac{\pi ^{2}}{6}\ln ^{2}2+\frac{1}{6}\ln ^{4}2+%
\frac{7}{2}\zeta (3)\ln 2+4Li_{4}(\frac{1}{2}),  \nonumber \\
C_{01} &=&\frac{2\ln \alpha }{\epsilon }-\ln ^{2}\alpha -4\ln \alpha
-2-\zeta (3),  \nonumber \\
C_{11} &=&\zeta (3),  \nonumber \\
C_{0} &=&\frac{\ln \alpha }{\epsilon }-\frac{\ln ^{2}\alpha }{2}-2\ln \alpha
-1-\zeta (3)-C_{0}^{^{\prime }},  \nonumber \\
C_{1} &=&4\ln 2\ln \alpha +2J_{1}(\alpha )-C_{0}^{^{\prime }}-2-\frac{\zeta
(3)}{2},  \nonumber \\
&-&4\ln 2-\frac{\pi ^{2}}{6}+4G,
\end{eqnarray}
where the Catalan constant $G=0.517\ldots $ appears as the integral

\[
G=-\int_{0}^{1}du\frac{\ln u}{1+u^2}. 
\]

\subsection{ The D-integrals}

\label{D}

\subsubsection{Definition}

These are integrals over the Feynman variables only. They originate from the
contractions which contain sums over both momenta and frequencies. 
\begin{equation}
D_{\nu }=\int []\frac{x_{3}^{\nu }(x_{1}+x_{2})^{\nu -1}(x_{i}x_{j})^{-\nu
-\epsilon }}{(\alpha x_{1}+x_{3})(\alpha x_{2}+x_{3})}.
\end{equation}

\subsubsection{$\epsilon$ expansion}

\begin{eqnarray}
D_{1} &=&-\ln ^{2}\alpha -\frac{\pi ^{2}}{6},  \nonumber \\
D_{2} &=&-2\ln \alpha .
\end{eqnarray}

\subsection{ The H - integrals}

\label{H}

\subsubsection{Definition}

The $H$-integrals involve the variable $z$ and the Feynman variables. All of
them originate from contractions with sums over both momenta and
frequencies. 
\begin{equation}
H_{\nu }=\int\limits_{\alpha }^{1}dz(z-\alpha )^{2\nu }\int []\frac{%
x_{2}^{2+\nu }(x_{1}+x_{3})(x_{i}x_{j})^{-2-\epsilon }}{(\alpha
x_{1}+zx_{2})(zx_{2}+x_{3})}.
\end{equation}

\subsubsection{$\epsilon$ expansion}

\begin{eqnarray}
H_{0} &=&-\ln \alpha +1,  \nonumber \\
H_{1} &=&-\ln \alpha .
\end{eqnarray}

\subsection{ The S - integrals}

\label{Ss}

\subsubsection{Definition}

These are integrals over the Feynman variables only and they do not not
contain the parameter $\alpha $. All of them originate from the expressions
with sums over both momenta and frequencies. 
\begin{equation}
S_{\mu \nu }=\int []\frac{x_{1}^{\mu }x_{2}^{1+\nu -\mu }((2-\nu -\mu
)x_{1}+x_{3})(x_{i}x_{j})^{-2-\epsilon }}{(x_{2}+x_{3})^{1+\nu }},
\end{equation}
\begin{equation}
S_{\nu }=\int [](x_{1}+x_{2})^{-1+2\nu }(x_{i}x_{j})^{-1-\nu -\epsilon }.
\end{equation}

\subsubsection{$\epsilon$ expansion}

\begin{eqnarray}
S_{00} &=&-\frac{1}{\epsilon }+2,  \nonumber \\
S_{01} &=&-\frac{1}{3\epsilon }+\frac{8}{9},  \nonumber \\
S_{11} &=&-\frac{1}{6\epsilon }+\frac{1}{9},  \nonumber \\
S_{0} &=&-\frac{1}{\epsilon }+2,  \nonumber \\
S_{1} &=&-\frac{2}{\epsilon }+2.
\end{eqnarray}

\subsection{The T-integrals}

\label{T}

\subsubsection{Definition}

The integrals are over the Feynman variables only. They come from the
expressions which only contain sums over frequency. 
\begin{equation}
T_{\mu \nu }^{\eta }=\frac{(1-\alpha )^{\eta }}{\alpha ^{\mu }}\int []\frac{%
x_{1}^{2-\eta }x_{2}^{\mu +\eta -1}x_{3}^{-1-\mu -\epsilon
}(x_{1}+x_{2})^{-2-\epsilon }}{(\alpha x_{2}+\nu \alpha x_{3}+x_{1})},
\end{equation}
\begin{equation}
T_{\mu \nu }=\int []\frac{x_{1}^{2\nu -2}(x_{1}+x_{2})^{-\nu -\epsilon
}(x_{1}+x_{3}+(\alpha +\mu )x_{2})}{x_{3}^{\nu +\epsilon }(\alpha
x_{2}+(1+\mu )x_{3}+x_{1})(x_{1}+x_{3}+\alpha x_{2})}.
\end{equation}

\subsubsection{$\epsilon$ expansion}

\begin{eqnarray}
T_{10}^{0} &=&-\frac{1}{\alpha }+1,  \nonumber \\
T_{11}^{0} &=&-\frac{1}{\alpha }+\frac{1}{\epsilon }+\ln \alpha +1, 
\nonumber \\
T_{20}^{0} &=&\frac{1}{6\alpha ^{2}}-\frac{1}{3\alpha }-\ln \alpha -\frac{11%
}{12},  \nonumber \\
T_{21}^{0} &=&\frac{1}{6\alpha ^{2}}+\frac{2}{3\alpha }+\frac{\ln \alpha +5/2%
}{\epsilon }+\frac{\ln ^{2}\alpha }{2}+4\ln \alpha +\frac{17}{12},  \nonumber
\\
T_{10}^{1} &=&-\frac{1}{\alpha }-2\ln \alpha -2,  \nonumber \\
T_{01} &=&\frac{\ln \alpha }{\epsilon }-\frac{\ln ^{2}\alpha }{2},  \nonumber
\\
T_{02} &=&\frac{1}{\epsilon },  \nonumber \\
T_{12} &=&-\frac{3\ln \alpha +11/2}{\epsilon }+\frac{3\ln ^{2}\alpha }{2}+%
\frac{9\ln \alpha }{2}  \nonumber \\
&-&4\ln 2\ln \alpha +\frac{\pi ^{2}}{6}-4Li_{2}(\frac{1}{2})-12\ln 2+\frac{27%
}{4}.
\end{eqnarray}

\subsection{List of symbols and abbreviations}

%
\begin{eqnarray}
\int []
&=&\int\limits_{0}^{1}dx_{1}\int\limits_{0}^{1}dx_{2}\int%
\limits_{0}^{1}dx_{3}\;\delta (x_{1}+x_{2}+x_{3}-1),  \nonumber \\
x_{ij} &=&x_{i}+x_{j},  \nonumber \\
x_{i}x_{j} &=&x_{1}x_{2}+x_{2}x_{3}+x_{3}x_{1}.
\end{eqnarray}
\begin{eqnarray}
K_{1}(\alpha ) &=&\int\limits_{\alpha }^{1}dz\int []\frac{%
x_{2}(x_{1}(x_{2}+x_{3})+x_{3}^{2})(x_{i}x_{j})^{-2-\epsilon }}{%
(zx_{2}+x_{3})(\alpha x_{1}+zx_{2}+2x_{3})},  \nonumber \\
J_{3}^{^{\prime }}(\alpha ) &=&\alpha \int\limits_{\alpha }^{1}\frac{dz}{z}%
\int []\frac{x_{2}(x_{1}(x_{2}+x_{3})+x_{3}^{2})(x_{i}x_{j})^{-2-\epsilon }}{%
(\alpha x_{1}+zx_{2}+2x_{3})^{2}},  \nonumber \\
J_{1}(\alpha ) &=&\int\limits_{\alpha }^{1}dz\int []\frac{%
x_{1}(x_{1}+x_{3})(x_{2}+x_{3})(x_{i}x_{j})^{-2-\epsilon }}{%
(zx_{1}+x_{3})(zx_{1}+\alpha x_{2}+2x_{3})},  \nonumber \\
C_{0}^{^{\prime }} &=&\int\limits_{\alpha }^{1}dzdy\int []\frac{%
x_{1}x_{2}^{2}(x_{i}x_{j})^{-2-\epsilon }}{%
(x_{3}+yx_{2})(zx_{1}+yx_{2}+2x_{3})}.
\end{eqnarray}
\setcounter{equation}{0}
\renewcommand{\theequation}{B.\arabic{equation}}

\section{Appendix B}

\label{AB}

In this appendix we present the calculation of the integral $A_{10}$ as a
typical example. We start with the integral 
\begin{eqnarray}
X_{10} &=&-\frac{32a^{3}}{\sigma _{0}d}\int\limits_{pq}p^{2}\sum_{k,m>0}m 
\nonumber \\
&&D_{p+q}^{c}(m)D^{3}D_{p}^{c}(k)D_{q}(k+m).
\end{eqnarray}
Using the Feynman trick, one can write 
\begin{eqnarray}
X_{10} &=&-\frac{16a^{3}}{\sigma _{0}d}\int\limits_{pq}p^{2}\int%
\limits_{0}^{\infty }dmm\int\limits_{0}^{\infty }dk  \nonumber \\
&&\Gamma (6)\int\limits_{\alpha }^{1}dz(z-\alpha )^{2}\int []\left[
{}\right. h_{0}^{2}+q^{2}x_{13}+p^{2}x_{12}+  \nonumber \\
&&2{\bf p}\cdot {\bf q}x_{1}+am(\alpha x_{1}+x_{3})+ak(zx_{2}+x_{3}\left.
{}\right] ^{-6}.
\end{eqnarray}
Shifting $q\rightarrow q-px_{1}/x_{13}$, we can decouple ${\bf p}$ and ${\bf %
q}$ in the denominator. We are then able to perform the integration over $%
k,m,p$ and $q$, resulting in 
\[
X_{10}=\frac{4\Omega _{d}^{2}h_{0}^{4\epsilon }}{\sigma _{0}\epsilon }%
A_{10}, 
\]
where 
\[
A_{10}=\int\limits_{\alpha }^{1}dz(z-\alpha )^{2}\int []\frac{%
x_{2}^{3}(x_{1}+x_{3})(x_{i}x_{j})^{-2-\epsilon }}{(zx_{2}+x_{3})(\alpha
x_{1}+x_{3})^{2}}. 
\]
Next we write the integral as a sum of four terms 
\begin{eqnarray}
A_{10} &=&\int\limits_{\alpha }^{1}\frac{dz(z-\alpha )^{2}}{z}\int []\frac{%
x_{2}(x_{1}+x_{3})(x_{i}x_{j})^{-1-\epsilon }}{(\alpha x_{1}+x_{3})^{2}} 
\nonumber \\
&\times &\left\{ {}\right. 1-x_{1}x_{3}(x_{i}x_{j})^{-1}-\frac{%
x_{3}(x_{1}+x_{3})(x_{i}x_{j})^{-1}}{z}  \nonumber \\
&+&\frac{x_{3}^{2}(x_{1}+x_{3})(x_{i}x_{j})^{-1}}{z(zx_{2}+x_{3})}\left.
{}\right\} =I_{0}-I_{1}-I_{2}+I_{3}.
\end{eqnarray}
In what follows we retain the full $\epsilon $ dependence in the $I_{0}$, $%
I_{1}$ and $I_{2}$ and it suffices to put $\epsilon =0$ in the fourth piece $%
I_{3}$. Introducing a change of variables 
\[
x_{1}=\frac{u}{s+1}\,;\,x_{2}=\frac{s}{s+1}\,;\,x_{3}=\frac{1-u}{s+1}, 
\]
where $0<s<\infty $ and $0<u<1$, then the four different pieces can be
written as follows 
\begin{eqnarray}
I_{0} &=&(\frac{1}{2}-2\alpha )\int\limits_{0}^{1}\frac{du}{(\alpha
u+1-u)^{2}}\int\limits_{0}^{\infty }ds\frac{s(s+1)^{2\epsilon }}{%
(s+u(1-u))^{1+\epsilon }},  \nonumber \\
I_{1} &=&\frac{1}{2}\int\limits_{0}^{1}du\frac{u(1-u)}{(\alpha u+1-u)^{2}}%
\int\limits_{0}^{\infty }ds\frac{s(s+1)^{2\epsilon }}{(s+u(1-u))^{2+\epsilon
}},  \nonumber \\
I_{2} &=&\int\limits_{0}^{1}du\frac{(1-u)}{(\alpha u+1-u)^{2}}%
\int\limits_{0}^{\infty }ds\frac{s(s+1)^{2\epsilon }}{(s+u(1-u))^{2+\epsilon
}},  \nonumber \\
I_{3} &=&\int\limits_{\alpha }^{1}dz\left( \frac{z-\alpha }{z}\right)
^{2}\int\limits_{0}^{1}du\frac{u(1-u)^{2}}{(\alpha u+1-u)^{2}}  \nonumber \\
&\times &\int\limits_{0}^{\infty }ds\frac{(s+1-u)}{(s+u(1-u))^{2}(\alpha
s+1-u)^{2}}.  \label{q}
\end{eqnarray}
The integrals over $s$ in Eq. (\ref{q}) can now be recognized as integral
representations of the hypergeometric function $_{2}F_{1}$. Write 
\begin{eqnarray}
I_{0} &=&(\frac{1}{2}-2\alpha )\int\limits_{0}^{1}du\frac{%
[u(1-u)]^{1-\epsilon }}{(\alpha u+1-u)^{2}}  \nonumber \\
&\times &\left[ {}\right. -\frac{1}{1+\epsilon }G_{0}(u(1-u))+\frac{1}{%
\epsilon }G_{1}(u(1-u))\left. {}\right] ,  \nonumber \\
I_{1} &=&\frac{1}{2}\int\limits_{0}^{1}du\frac{[u(1-u)]^{1-\epsilon }}{%
(\alpha u+1-u)^{2}}  \nonumber \\
&\times &\left[ {}\right. -\frac{1}{\epsilon }G_{1}(u(1-u))-\frac{1}{%
1-\epsilon }G_{2}(u(1-u))\left. {}\right] ,  \nonumber \\
I_{2} &=&\int\limits_{0}^{1}du\frac{u^{-\epsilon }(1-u)^{1-\epsilon }}{%
(\alpha u+1-u)^{2}}  \nonumber \\
&\times &\left[ {}\right. -\frac{1}{\epsilon }G_{1}(u(1-u))-\frac{1}{%
1-\epsilon }G_{2}(u(1-u))\left. {}\right] ,  \nonumber \\
I_{3} &=&\int\limits_{\alpha }^{1}dz\left( \frac{z-\alpha }{z}\right)
^{2}\int\limits_{0}^{1}du\frac{u}{zu+1-u}  \nonumber \\
&\times &\left[ {}\right. \frac{1}{2}H_{3}(1-\alpha u)+\frac{(1-u)}{u}\frac{%
\Gamma (3)}{\Gamma (4)}H_{4}(1-\alpha u)\left. {}\right] ,
\end{eqnarray}
then, in the limit where $\epsilon \rightarrow 0$, we can identify the
functions $G_{i}$ and $H$ as follows 
\begin{eqnarray}
G_{0}(1-z) &=&{_{2}F_{1}}(1,-2\epsilon ,-\epsilon ;z)=\frac{1+z}{1-z}, 
\nonumber \\
G_{1}(1-z) &=&{_{2}F_{1}}(1,-2\epsilon ,1-\epsilon ;z)=1+2\epsilon \ln (1-z),
\nonumber
\end{eqnarray}
\begin{eqnarray}
G_{2}(1-z) &=&{_{2}F_{1}}(1,-2\epsilon ,2-\epsilon ;z)=1,  \nonumber \\
H_{3}(z) &=&{_{2}F_{1}}(1,2,3;z)=-\frac{2}{z^{2}}\left( \ln (1-z)+z\right) ,
\nonumber \\
H_{4}(z) &=&{_{2}F_{1}}(1,2,4;z)=\frac{6}{z^{3}}\left( {}\right. (1-z)\ln
(1-z)  \nonumber \\
&+&z-z^{2}/2\left. {}\right) .
\end{eqnarray}
Using these results we obtain 
\begin{eqnarray}
I_{0} &=&-\frac{1}{\alpha }-\frac{\ln \alpha +2}{\epsilon }-\frac{\ln
^{2}\alpha }{2}-2\ln \alpha -\frac{\pi ^{2}}{3},  \nonumber \\
I_{1} &=&\frac{\ln \alpha +2}{\epsilon }+\frac{\ln ^{2}\alpha }{2}+2\ln
\alpha +\frac{\pi ^{2}}{3},  \nonumber \\
I_{2} &=&\frac{\ln \alpha +1}{\epsilon }+\frac{\ln ^{2}\alpha }{2}+2\ln
\alpha +\frac{\pi ^{2}}{3}+1,  \nonumber \\
I_{3} &=&-\ln \alpha .
\end{eqnarray}
The final answer is therefore 
\begin{eqnarray}
A_{10} &=&-\frac{1}{\alpha }-\frac{2\ln \alpha +3}{\epsilon }-\ln ^{2}\alpha
-5\ln \alpha  \nonumber \\
&-&\frac{2\pi ^{2}}{3}+3
\end{eqnarray}

\end{multicols}

\end{document}